\documentclass[10pt]{article}%
\usepackage[a4paper,top=2.8cm,bottom=2.8cm,left=3cm,right=3cm]{geometry} %
\usepackage{sgame}
\usepackage{amsmath}%
\usepackage{amsthm}%
\usepackage{amsfonts}%
\usepackage{amssymb}%
\usepackage{mathrsfs}
\usepackage{braket}					
\usepackage{mathtools}
\usepackage{enumitem}
\usepackage{xcolor}%
\usepackage{graphicx}%
\usepackage{bm}
\usepackage{tikz}%
\usetikzlibrary{calc}%
\usetikzlibrary{cd}%
\usepackage{float}%
\usepackage{etoolbox}
\usepackage{changepage}
\usepackage{datetime}
\makeatletter
\newcommand\Romanmonth{\@Roman{\month}}
\makeatother
\newdateformat{mydate}{\THEDAY.\Romanmonth.\THEYEAR}
\usepackage{natbib}          
\usepackage[T1]{fontenc}
\usepackage[utf8]{inputenc}
\usepackage{epigraph}
\makeatletter
\patchcmd{\epigraph}{\@epitext{#1}}{\itshape\@epitext{#1}}{}{}
\makeatother
\usepackage{titlesec}
\titleformat{\section}
  {\normalfont\sc\centering}
  {\thesection.}{1em}{}
\titleformat{\subsection}
  {\normalfont\centering}
  {\normalfont\sc\centering\thesubsection}{1em}{}{}
\titleformat{\subsubsection}
  {\normalfont\slshape\centering}
  {\normalfont\sc\centering\thesubsubsection}{1em}{}{}
\usepackage{xparse}
\AtBeginDocument

\DeclareFontFamily{U}{ntxmia}{}
\DeclareFontShape{U}{ntxmia}{m}{it}{<-> ntxmia }{}
\DeclareFontShape{U}{ntxmia}{b}{it}{<-> ntxbmia }{}
\DeclareSymbolFont{lettersA}{U}{ntxmia}{m}{it}
\SetSymbolFont{lettersA}{bold}{U}{ntxmia}{b}{it}

\ExplSyntaxOn
\NewDocumentCommand{\varmathbb}{m}
 {
  \tl_map_inline:nn { #1 }
   {
    \use:c { varbb##1 }
   }
 }
\tl_map_inline:nn { ABCDEFGHIJKLMNOPQRSTUVWXYZ }
 {
  \exp_args:Nc \DeclareMathSymbol{varbb#1}{\mathord}{lettersA}{\int_eval:n { `#1+64 }}
 }
\exp_args:Nc \DeclareMathSymbol{varbbk}{\mathord}{lettersA}{169}
\ExplSyntaxOff
\makeatletter
\g@addto@macro\@floatboxreset\centering
\makeatother
\makeatletter
\def\bbordermatrix#1{\begingroup \m@th
  \@tempdima 4.75\p@
  \setbox\z@\vbox{%
    \def\cr{\crcr\noalign{\kern2\p@\global\let\cr\endline}}%
    \ialign{$##$\hfil\kern2\p@\kern\@tempdima&\thinspace\hfil$##$\hfil
      &&\quad\hfil$##$\hfil\crcr
      \omit\strut\hfil\crcr\noalign{\kern-\baselineskip}%
      #1\crcr\omit\strut\cr}}%
  \setbox\tw@\vbox{\unvcopy\z@\global\setbox\@ne\lastbox}%
  \setbox\tw@\hbox{\unhbox\@ne\unskip\global\setbox\@ne\lastbox}%
  \setbox\tw@\hbox{$\kern\wd\@ne\kern-\@tempdima\left[\kern-\wd\@ne
    \global\setbox\@ne\vbox{\box\@ne\kern2\p@}%
    \vcenter{\kern-\ht\@ne\unvbox\z@\kern-\baselineskip}\,\right]$}%
  \null\;\vbox{\kern\ht\@ne\box\tw@}\endgroup}
\makeatother
\makeatletter
\newcommand{\mathleft}{\@fleqntrue}
\newcommand{\mathcenter}{\@fleqnfalse}
\makeatother
\DeclareMathOperator{\supp}{supp}
\DeclareMathOperator{\proj}{proj}

\DeclareMathOperator{\K}{\mbb{K}}

\DeclareMathOperator{\Bel}{\mbb{B}}
\DeclareMathOperator{\CB}{\mbb{CB}}

\DeclareMathOperator*{\argmax}{arg\max}

\usepackage{scalerel}

\makeatletter
\def\th@plain{%
  \thm@notefont{}
  \itshape 
}
\def\th@definition{%
  \thm@notefont{}
  \normalfont 
}
\makeatother
\numberwithin{equation}{section}
\usepackage{stmaryrd}
\newcommand{\evaluation}[2][]{\ensuremath{\llbracket #2\rrbracket_{#1}}}
\newcommand{\mscr}{\mathscr}
\newcommand{\mbf}{\mathbf}
\newcommand{\mbb}{\mathbb}
\newcommand{\mfrak}{\mathfrak}
\newcommand{\bN}{\mathbb{N}}

\newcommand{\Llra}{\Longleftrightarrow}

\newcommand{\la}{\langle}
\newcommand{\ra}{\rangle}
\newcommand{\Ad}{\mathsf{A}}
\newcommand{\PR}{\mbf{PR}}
\newcommand{\WR}{\mbf{WR}}
\newcommand{\BR}{\mbf{BR}}
\newcommand{\TR}{\mbf{R}}

\newcommand{\AR}{\mbf{SR}}
\newcommand{\YR}{\mbf{YR}}
\def\DR{\futurelet\next\DRx}
\def\DRx{\ifx^\next\afterfi{\DRy}\else\afterfi{\DRy^{}}\fi}
\def\DRy^#1{\mbf{R}^{\Delta, #1}}
\def\afterfi#1#2\fi{\fi#1}
\newcommand{\obr}{\rho^{\max}}
\newcommand{\pbr}{\rho^{\min}}
\newcommand{\bbr}{\rho^{\mbf{A}}}

\newcommand{\Pes}{\mathsf{P}}
\newcommand{\Opt}{\mathsf{O}}
\newcommand{\Att}{\mathsf{E}}
\newcommand{\Mar}{\mathsf{M}}
\newcommand{\Deg}{\mathsf{D}}

\newcommand{\PCBP}{\mathsf{PCBP}}
\newcommand{\OCBO}{\mathsf{OCBO}}

\newcommand{\fob}{\varphi}

\newcommand{\ACBA}{\mathsf{ACBA}}

\providecommand{\envert}[2][-1]{
\ensuremath{\mathinner{
\ifthenelse{\equal{#1}{-1}}{ 
\!\left\lvert#2\right\rvert}{}
\ifthenelse{\equal{#1}{0}}{ 
\lvert#2\rvert}{}
\ifthenelse{\equal{#1}{1}}{ 
\!\bigl\lvert#2\bigr\rvert}{}
\ifthenelse{\equal{#1}{2}}{ 
\!\Bigl\lvert#2\Bigr\rvert}{}
\ifthenelse{\equal{#1}{3}}{ 
\!\biggl\lvert#2\biggr\rvert}{}
\ifthenelse{\equal{#1}{4}}{ 
\!\Biggl\lvert#2\Biggr\rvert}{}
}} 
}
\let\abs=\envert
\usepackage{mleftright}

{\catcode`\|=\active
  \xdef\set{\protect\expandafter\noexpand\csname sets \endcsname}
  \expandafter\gdef\csname sets \endcsname#1{\mathinner
        {\lbrace\,{\mathcode`\|32768\let|\midvert #1}\,\rbrace}}
  \xdef\Sets{\protect\expandafter\noexpand\csname Sets \endcsname}
  \expandafter\gdef\csname Sets \endcsname#1{\left\{
     	\ifx\SavedDoubleVert\relax \let\SavedDoubleVert\|\fi
     	{\let\|\SetDoubleVert
     	\mathcode`\|32768\let|\SetsVert
     	#1}
     	\right\}}
}

\def\midvert{\egroup\mid\bgroup}
\def\SetsVert{\@ifnextchar|{\|\@gobble}
    	{\egroup
    	\;
    	\mid@vertical
    	\;
    	\bgroup}}
\def\SetsDoubleVert{\egroup
	\;
	\mid@dblvertical
	\;
	\bgroup}

{\catcode`\|=\active
  \xdef\Round{\protect\expandafter\noexpand\csname Round \endcsname}
  \expandafter\gdef\csname Round \endcsname#1{\left( %
     \ifx\SavedDoubleVert\relax \let\SavedDoubleVert\|\fi
     \:{\let\|\RoundDoubleVert
     \mathcode`\|32768\let|\RoundVert
     #1}\:\right)}}

\def\midvert{\egroup\mid\bgroup}
\def\RoundVert{\@ifnextchar|{\|\@gobble}
    {\egroup\:\mid@vertical\:\bgroup}}
\def\RoundDoubleVert{\egroup\:\mid@dblvertical\:\bgroup}

{\catcode`\|=\active
  \xdef\Rounds{\protect\expandafter\noexpand\csname Rounds \endcsname}
  \expandafter\gdef\csname Rounds \endcsname#1{\mleft(
     	\ifx\SavedDoubleVert\relax \let\SavedDoubleVert\|\fi
     	{\let\|\RoundsDoubleVert
     	\mathcode`\|32768\let|\RoundsVert
     	#1}
     	\mright)}}

\def\midvert{\egroup\mid\bgroup}
\def\RoundsVert{\@ifnextchar|{\|\@gobble}
  	 {\egroup%
   	 \hspace{0.01cm}
   	 \mid@vertical
    	\hspace{0.01cm}
    	\bgroup}}
\def\RoundsDoubleVert{\egroup
	\hspace{0.01cm}
	\mid@dblvertical
	\hspace{0.01cm}
	\bgroup}

{\catcode`\|=\active
  \xdef\Square{\protect\expandafter\noexpand\csname Square \endcsname}
  \expandafter\gdef\csname Square \endcsname#1{\left[ %
     \ifx\SavedDoubleVert\relax \let\SavedDoubleVert\|\fi
     \:{\let\|\SquareDoubleVert
     \mathcode`\|32768\let|\SquareVert
     #1}\:\right]}}

\def\midvert{\egroup\mid\bgroup}
\def\SquareVert{\@ifnextchar|{\|\@gobble}
    {\egroup\:\mid@vertical\:\bgroup}}
\def\SquareDoubleVert{\egroup\:\mid@dblvertical\:\bgroup}

{\catcode`\|=\active
  \xdef\Squares{\protect\expandafter\noexpand\csname Squares \endcsname}
  \expandafter\gdef\csname Squares \endcsname#1{\mleft[
     	\ifx\SavedDoubleVert\relax \let\SavedDoubleVert\|\fi
     	{\let\|\SquaresDoubleVert
     	\mathcode`\|32768\let|\SquaresVert
     	#1}
     	\mright]}}

\def\midvert{\egroup\mid\bgroup}
\def\SquaresVert{\@ifnextchar|{\|\@gobble}
  	 {\egroup%
   	 \hspace{0.01cm}
   	 \mid@vertical
    	\hspace{0.01cm}
    	\bgroup}}
\def\SquaresDoubleVert{\egroup
	\hspace{0.01cm}
	\mid@dblvertical
	\hspace{0.01cm}
	\bgroup}
\usepackage{varioref}
\usepackage[pagebackref,hypertexnames=false]{hyperref}
\usepackage[noabbrev,nameinlink]{cleveref}
\definecolor{egtgreen}{RGB}{75, 155, 8}
\definecolor{egtpurple}{RGB}{10, 33, 128}
\definecolor{egtred}{RGB}{103, 25, 15}
\definecolor{burgundy}{rgb}{0.5, 0.0, 0.13}
\definecolor{cyanp}{RGB}{45,129,173}
\hypersetup{colorlinks=true,linkcolor={egtred},citecolor=green!50!black,bookmarksnumbered=true}
\hypersetup{%
pdfcreator={},
  pdftitle={Optimism and Pessimism in Strategic Interactions under Ignorance},%
  pdfauthor={},%
  pdfsubject={},%
  pdfproducer={},%
  pdfkeywords={}]%
}
\newcommand{\Mref}[2][cyanp]{%
\hypersetup{linkcolor=#1}%
\Cref{#2}%
\hypersetup{linkcolor=burgundy}%
}
\newcommand{\Sref}[2][burgundy]{%
\hypersetup{linkcolor=#1}%
\Cref{#2}%
\hypersetup{linkcolor=burgundy}%
}
\usepackage{xpatch}\xpatchbibmacro{pageref}{parens}{brackets}{}{}
\makeatletter
\patchcmd{\BR@backref}{\newblock}{\newblock[}{}{}
\patchcmd{\BR@backref}{\par}{]\par}{}{}
\makeatother
\theoremstyle{plain}
\newtheorem{theorem}{Theorem}
\newtheorem{definition}{Definition}[section]
\newtheorem{proposition}[theorem]{Proposition}
\newtheorem{lemma}{Lemma}
\newtheorem{corollary}[theorem]{Corollary}

\newtheorem{remark}{Remark}[section]
\theoremstyle{definition}
\theoremstyle{remark}
\makeatletter
\def\thm@space@setup{%
  \thm@preskip=0.5cm 
  \thm@postskip=\thm@preskip 
}
\makeatother
\newtheoremstyle{myclaimstyle} 
    {\topsep}                    
    {\topsep}                    
    {\itshape}                   
    {}                           
    {\itshape}                   
    {.}                          
    {.5em}                       
    {}  

\theoremstyle{myclaimstyle}

\newtheoremstyle{mycasestyle} 
    {3pt}                    
    {3pt}                    
    {}                   
    {}                           
    {}                   
    {:}                          
    {.5em}                       
    {}  

\theoremstyle{mycasestyle}

\newtheoremstyle{named}{}{}{\itshape}{}{\bfseries}{.}{.5em}{\thmnote{#3}}
\theoremstyle{named}

\makeatletter
\newtheoremstyle{case}
  {5pt}
  {5pt}
  {\addtolength{\@totalleftmargin}{0em}
   \addtolength{\linewidth}{-1em}
   \parshape 1 1em \linewidth}
  {}
  {\normalfont}
  {:}
  {.5em}
  {}
\makeatother

\theoremstyle{case}

\AtEndEnvironment{proof}{\setcounter{step}{0}}
\AtEndEnvironment{proof}{\setcounter{claim}{0}}
\AtEndEnvironment{proof}{\setcounter{case}{0}}
\AtEndEnvironment{proof}{\setcounter{Cases}{0}}
\makeatletter
\def\thm@space@setup{%
  \thm@preskip=0.5cm 
  \thm@postskip=\thm@preskip 
}
\makeatother
\usepackage{thmtools}

\declaretheorem[style=definition, name=Example, qed=$\diamond$]{example}
\renewcommand\thmcontinues[1]{Continued}
\declaretheoremstyle[
  spaceabove=3pt, spacebelow=6pt,
  headfont=\normalfont\itshape,
  notefont=\mdseries, notebraces={(}{)},
  bodyfont=\normalfont,
  postheadspace=1em
]{innerproof}

\renewcommand\thmcontinues[1]{}

\renewcommand\thmcontinues[1]{}
\usepackage{footnote}
\makesavenoteenv{minipage}

\usetikzlibrary{positioning,shapes.callouts}
\newcommand{\GG}[1]{}
\title{\vspace{-1.3cm}Optimism and Pessimism in Strategic Interactions under Ignorance%
\thanks{The authors would like to thank Emiliano Catonini, Francesco De Sinopoli, Amanda Friedenberg, Michele Gori, Michael Greinecker, Cristoph Kuzmics, Burkhard Schipper, Peio Zuazo-Garin, the anonymous associate editor who handled the paper, various anonymous referees, the audiences of the special session on economic theory of the NOeG 2021 conference, of the GRASSXV workshop, of the CEPET2022 workshop, of the $2^{\text{nd}}$ Durham Economic Theory Conference, of the LOFT2022 Conference, and of the seminars at the Department of Economics at the University of Verona, the Queen's Management School at Queen's University Belfast, and the Department of Business Decisions and Analytics at the University of Vienna. Of course, all errors are our own. Previous versions of this paper circulated under various titles. For the purpose of open access, the authors have applied a `Creative Commons Attribution (CC BY) licence' to any Author Accepted Manuscript version arising from this submission. Pierfrancesco thankfully acknowledges financial support from the Austrian Science Fund (FWF) (P31248-G27) and from MIUR under the PRIN 2017 program (grant number 2017K8ANN4).}
\\
}
\author{%
Pierfrancesco Guarino\thanks{%
University of Udine (Department of Economics and Statistics -- DIES). %
\textit{E-mail:} \texttt{pf.guarino@hotmail.com} \& \texttt{pierfrancesco.guarino@uniud.it}.} 
{\ \& }%
Gabriel Ziegler\thanks{%
 	University of Edinburgh (School of Economics) \& University of Pittsburgh (Department of Economics). %
 	\textit{E-mail:} \texttt{ziegler@ed.ac.uk}.}%
}
\date{\mydate\today}
\begin{document}
\renewcommand\thmcontinues[1]{Continued}

\maketitle

\vspace{-0.5cm}
\begin{center}
FINAL VERSION\\\vspace{0.1cm}
\textit{Accepted for Publication on ``Games and Economic Behavior''}
\end{center}
\vspace{0.3cm}


\begin{abstract}
\noindent %
We study players interacting under the veil of ignorance, who have---coarse---beliefs represented as subsets of opponents’ actions. We analyze when these players follow $\max \min$ or $\max\max$ decision criteria, which we identify with pessimistic or optimistic attitudes, respectively. Explicitly formalizing these attitudes and how players reason interactively under ignorance, we characterize the behavioral implications related to common belief in these events: while optimism is related to Point Rationalizability, a new algorithm---Wald Rationalizability---captures pessimism. Our characterizations allow us to uncover novel results: ($i$) regarding optimism, we relate it to wishful thinking  \emph{\'a la} \cite{Yildiz_2007} and we prove that dropping the (implicit) ``belief-implies-truth'' assumption  reverses an existence failure described therein; ($ii$) we shed light on the notion of rationality in ordinal games; ($iii$) we clarify the conceptual underpinnings behind a discontinuity in Rationalizability hinted in the analysis of \citet{Weinstein_2016}.

\bigskip



\end{abstract}


\section{Introduction}
\label{sec:introduction}

\subsection{Motivation \& Results}
\label{subsec:motivation_results}

Games played under the veil of \emph{ignorance}\footnote{See \cite{Milnor_1954},  \citet[Chapter 13]{Luce_Raiffa_1957}, \cite{Arrow_Hurwicz_1977}, and \cite{Kelsey_Quiggin_1992}.} are all those strategic interactions where players are not able to describe their beliefs via any form of probability measure (and the like). This essentially implies that these strategic interactions lie outside the realm of the Bayesian paradigm \emph{{\`a} la} \cite{Savage_1954} or the models on ambiguity  stemming from \cite{Schmeidler_1989} and \cite{Gilboa_Schmeidler_1989}.\footnote{See the comprehensive survey \cite{Gilboa_Marinacci_2011}.}  As a result, whenever we, as analysts, study games under ignorance, we lack all those---nice---mathematical properties coming from the assumption of working with cardinal utility that, starting from \cite{vonNeumann_Morgenstern_1944} and \cite{Nash_1950,Nash_1951},\footnote{Even some of the earliest contributions to game theory, e.g., \cite{Borel_1921, Borel_1927} and \cite{vonNeumann_1928},  relied on these nice properties to establish the famous minimax theorem.} considerably generalize the scope of game theory: indeed, lacking probability measures (and the like) essentially results in not being able to perform  expected utility computations.

In light of this, it seems natural to ask why there should be any interest in games under ignorance, where players' beliefs are---rather coarsely---represented simply via collections of their opponents' actions. Indeed, one might wonder if there is any hope of obtaining reasonable predictions for strategic interactions under ignorance when these nice mathematical structures are not applicable. Bypassing the---given the appropriate circumstances---descriptive accuracy provided by assuming to be in presence of a strategic interaction under ignorance, we show that tight predictions are still possible even though the assumptions underlying our analysis are naturally rather weak. Indeed, since probability measures play no role and---as a result---risk attitudes are out of place, the study of games under ignorance relies essentially on the framework provided by ordinal games. That is, in our analysis only ordinal preferences over the outcomes of the game are transparent between the players.\footnote{See \cite{Bonanno_2018} for a textbook focusing on this \emph{variety} of games. It has to be observed that, throughout this work, we use the word ``variety'' to refer to games played under ignorance rather than the word ``class''. Indeed, assuming the presence of ignorance does not have any impact on the actual description of the primitives of the games, which is what should be effected in order to distinguish between different classes of games.} This in turn implies that the predictions we obtain in this context cover all those strategic interactions where we, as analysts, simply assume to have knowledge on our side of the \emph{ordinal} preferences of the players and their transparency\footnote{That is, common knowledge in the informal sense of the expression.} between them, along with the characterizing assumption that  players' beliefs are coarsely represented via collections of their opponents' actions. As such, this endeavor can be considered in line with the so called \emph{Wilson doctrine} going back to \cite{Wilson_1987}, that asks for a relaxation of the common knowledge assumptions in game theory.\footnote{With the understanding that---implicitly---our analysis relies on assuming common knowledge of a situation under ignorance.}

By focusing on games under ignorance, any strategic analysis immediately runs into a conceptual issue. Whereas there is general agreement on the notion of rationality---whenever explicitly posited---that should be employed in the analysis of `standard' games,\footnote{That is, games where players have cardinal utilities and have beliefs in the form of probability measures.} namely, that of Bayesian rationality as subjective expected utility maximization, the same cannot be stated regarding games under ignorance---as ordinal games. We bypass this conceptual issue by focusing on two classic decision criteria under ignorance, namely, $\max \max$ and $\max \min$, building on their decision-theoretic foundation going back to \cite{Wald_1950}, \cite{Milnor_1954}, and  \cite{Arrow_Hurwicz_1977}.\footnote{The $\max \min$ decision criterion goes back to \citet[Chapter 1.4.2, p.18]{Wald_1950}. The $\max \max$ criterion can be obtained by replacing the convexity axiom in \cite{Milnor_1954} with a concavity axiom.}   Thus, informally, an ignorant  player---with her  rather coarse beliefs---can have the following different attitudes concerning her play:
\begin{itemize}
	\item she can be \emph{optimistic}, in which case she is going to assume that, for every action of hers, her opponents choose their actions (from the actions she contemplates as possible), to \emph{maximize} her utility and---consequently---she is going to choose the action that gives her the highest utility accordingly (hence, proceeding according to the $\max \max$ criterion);
	
	\item she can be \emph{pessimistic}, in which case she is going to assume that, for every action of hers, her opponents choose their actions (from the actions she contemplates as possible), to \emph{minimize} her utility and---consequently---she is going to choose the action that gives her the highest utility accordingly (thus, proceeding according to the $\max \min$ criterion).
\end{itemize}

Moving from those decision criteria, we study what are the behavioral implications that we (or the players themselves) can expect to obtain when we (or they) assume that players are optimistic (resp., pessimistic), that there is optimism and mutual belief in optimism (resp., pessimism), and so on up to optimism and common belief in optimism (resp., pessimism). Thus, to provide an explicit analysis of strategic reasoning under ignorance, we perform our investigation by employing the tools of epistemic game theory. With respect to this point, first of all, we identify in \Mref{def:epistemic_possibility_structure} the framework appropriate for our analysis, which happens to be that of \emph{epistemic possibility structures}  of \citet[Sections 2 \& 3]{Mariotti_et_al_2005}. The main difference between epistemic possibility structures and (the more commonly used) epistemic type structures\footnote{See \cite{Dekel_Siniscalchi_2015} for a comprehensive survey  of epistemic game theory or \cite{Perea_2012} for a textbook completely devoted to the topic.} is that in epistemic possibility structures beliefs are represented exactly in the coarse way described above, i.e., as subsets of the space of uncertainty.

With epistemic possibility structures at our disposal, we proceed by explicitly defining those epistemic events that correspond to a player being optimistic or pessimistic (respectively, in \Mref{eq:optimism} and \Mref{eq:pessimism}), thus formalizing the---informal---description of these attitudes provided above. Also, by employing modal operators capturing belief and common belief (as it is standard in the epistemic game theory literature), we define in \Mref{def:ECBE} the events in epistemic possibility structures of \emph{Optimism/Pessimism and Common Belief in Optimism/Pessimism}. Having established the epistemic events of interest, we proceed by providing a characterization of their behavioral implications. In \Mref{def:point_r}, we give a definition in our language of Point Rationalizability of \cite{Bernheim_1984}, that algorithmically characterizes the behavior under \emph{Optimism and Common Belief in Optimism} (as established in \Mref{th:tropical_p}). Furthermore, in \Mref{def:vn_r}, we introduce a new procedure with a Rationalizability flavor, called \emph{Wald Rationalizability}, that---as we show in \Mref{th:tropical_vn}---algorithmically characterizes the behavioral implications of \emph{Pessimism and Common Belief in Pessimism}.  Regarding a comparison of the two algorithmic procedures, \Mref{prop:P-V} shows that the set of actions that survive Point Rationalizability is a subset of the set of actions that survive Wald Rationalizability. In other words, Pessimism and Common Belief in Pessimism has less identification power through observed behavior than its optimistic counterpart.

Taking into accounts the peculiarities of working with epistemic possibility structures, our characterization theorems follow arguments that are now common in the epistemic game theory literature. Nevertheless, there is an interesting detail in \Mref{th:tropical_p} and \Mref{th:tropical_vn}, namely, our language allows us to provide \emph{one} proof for the \emph{two} characterizations (as in \Sref{subsec:proofs_algo}). However, we do see the main contribution of the present paper---obviously, beyond the introduction of the $\max \max$ and $\max \min$ decision criteria in a game theoretical context---to be conceptual in providing a precise language to address questions for games under ignorance. To appreciate this last statement, we use our main characterizations to shed light on some existing---and seemingly unrelated---results.

First of all, in \Sref{sec:Yildiz_2007} we compare our notion of Optimism to the notion of Wishful Thinking  introduced in \cite{Yildiz_2007}. Quite naturally, our Optimism and Common Belief in Optimism is essentially equivalent to  Wishful Thinking and Common Belief in Wishful Thinking. However, we illustrate that the `wishful thinking'-oriented algorithm of \citet[Section 3, p.326]{Yildiz_2007} crucially, but somewhat implicitly, relies on \emph{knowledge} instead of \emph{belief}. It is well known that the key difference between these modal attitudes is the \emph{Truth axiom}, which asserts that whatever a payer believes is also true.  In the realm of  static `standard' games the belief-knowledge distinction is often inconsequential,\footnote{See \Sref{subsec:CB_CK} for a more detailed discussion on this point. For dynamic games it is well known that the distinction might matter, as argued in \cite{Samet_2013}.} but this is not the case for games under ignorance. An important consequence of this conceptual remark is that Wishful Thinking and Common Belief in Wishful Thinking is never empty: an important property that is orthogonal to the existence failure of Wishful Thinking and Common \emph{Knowledge} in Wishful Thinking as established in \citet[Example 5, pp.334--335]{Yildiz_2007}. Furthermore, our analysis illustrates that, in contrast to \cite{Yildiz_2007}, who employs probability measures and (transparency of) players' risk attitudes, the rather weak assumptions described above are sufficient to study wishful thinking in strategic environments.

In \Sref{sec:B-dominance}, rather naturally given that we work in an `ordinal' setting, we investigate the relation between our notions of Rationalizability and a form of Rationalizability built on B{\"o}rgers Dominance, in light of the latter being a notion specifically designed for ordinal games. Thus, first of all, it has to be recalled that B{\"o}rgers Dominance has been introduced\footnote{As ``Pure Strategy Dominance'' in the title, whereas in the body of the article it is simply called ``dominance''.} in \citet[Definition 4, p.426]{Borgers_1993} to capture in ordinal games the notion  of \emph{justifiability},\footnote{\cite{Borgers_1993} actually does not use the word ``justifiability'': rather, he calls an action \emph{rational} if it satisfies the condition described in the main body. The recent literature on epistemic game theory distinguishes rationality and justifiability in the following way: an action is justifiable, while an \emph{action-type pair} is rational  (see \cite{Battigalli_et_al_forthcoming}). Since our contribution is related to the epistemic game theory literature, we employ this recent terminology.} which in the original article means that for every justifiable action we can produce a probability measure and a von Neumann-Morgenstern utility function that agrees with the player's ordinal preferences according to which the action is a maximizer. Thus, as such, this dominance notion  is---by definition---directly linked to the standard realm of the usual Bayesian paradigm. Given this, starting from a notion of \emph{rationality} defined as choosing an action that is weakly undominated by a pure action relative to the opponents' actions that are deemed possible,\footnote{See \Sref{sec:B-dominance} and \Sref{subsec:B_Rationality} for a thorough discussion of this notion of rationality.} \citet[Theorem 1, p.5]{Bonanno_Tsakas_2018} show that Rationality and Common Belief in Rationality (as defined above) is algorithmically characterized by an algorithmic procedure that iteratively eliminates actions that are B{\"o}rgers dominated. To be able to compare our notions with the one in \cite{Bonanno_Tsakas_2018}, in \Mref{eq:rationality} we define in our language the notion of rationality of \citet[Definition 2, p.4]{Bonanno_Tsakas_2018}, that we---rather naturally given its definition---call ``Admissibility'',\footnote{Where we define it for pure actions only as in \citet[Chapter 13.3, p.287]{Luce_Raiffa_1957}. \citet[Definition 3.1, p.320]{Brandenburger_et_al_2008} is the corresponding version in presence of mixed actions.}  and we replicate \citet[Theorem 1, p.5]{Bonanno_Tsakas_2018} in our setting, where the focus is on the players' perspective rather than on that of an outside analyst, by showing that \emph{Admissibility and Common Belief in Admissibility}  is algorithmically characterized by an appropriately defined version of B{\"o}rgers Rationalizability (as in \Mref{def:IBD}). Armed with this result as a benchmark, we compare B{\"o}rgers Rationalizability to Point Rationalizability and Wald Rationalizability: while we can state in \Mref{prop:P_B}  that Point Rationalizability always selects a subset of the profiles of actions selected by B{\"o}rgers Rationalizability, we show that there is no inclusion relation between B{\"o}rgers and Wald Rationalizability. Thus, for ordinal games, the predictions based on a clear decision-theoretic optimality criterion for situations under ignorance might be distinct from those obtained when the baseline assumption of behavior is derived either from a Bayesian notion or, equivalently, from a particular dominance notion. However, interestingly, we establish in \Mref{prop:generic_V_B} that, in generic games, B{\"o}rgers Rationalizability always selects a superset of the profiles of actions selected by Wald Rationalizability. Therefore, we can conclude that, in generic ordinal games, common belief in either optimism or pessimism refines B{\"o}rgers Rationalizability. That is, our two decision-theoretic notions under ignorance provide sharper predictions than the one based on the Bayesian approach.

Finally, in \Sref{sec:rationalizability}, we focus on the relation between the present setting and that of \cite{Weinstein_2016}, a recent important contribution that---among the other things---studies the behavior of the Rationalizability correspondence as players become infinitely risk averse or risk seeking, considering the payoffs of a given game as \emph{monetary payoffs}. As a matter of fact, this section is actually related to \Sref{sec:B-dominance}, since here as well B{\"o}rgers Rationalizability happens to enter the stage. Indeed, \citet[Proposition 3, p.1888]{Weinstein_2016} shows that the set of (standard) rationalizable action profiles converges to the---opportunely defined---B{\"o}rgers rationalizable action profiles as players become infinitely risk averse, whereas point rationalizable action profiles are the result of players becoming infinitely risk seeking (as shown in \citet[Proposition 2, p.1887]{Weinstein_2016}). Focusing on the corresponding limit points, while Point Rationalizability is the limit point of players being infinitely risk seeking, B{\"o}rgers Rationalizability does \emph{not} coincide with the limit point of players being infinitely risk averse. This actually corresponds to the discontinuity hinted in \citet[p.1891]{Weinstein_2016}. Indeed, it is  Wald Rationalizability that does coincide with the limit point of players being infinitely risk averse. Thus, by identifying this along with the epistemic characterization of Wald Rationalizability given in \Mref{th:tropical_vn}, we provide a conceptual foundation to this phenomenon. As a matter of fact, this hides an interesting conceptual twist, namely that the discontinuity along with the epistemic characterization seem to be a symptom of the fact that, even if defined for ordinal games, B{\"o}rgers dominance is fundamentally related to the standard Bayesian framework, a point which is consistent with the \emph{rationale} behind this notion of dominance.

\subsection{Related Literature}
\label{subsec:related_literature}

This paper fits various streams of literature. On one side, it belongs to those studies that focus on games where only \emph{ordinal preferences} are assumed to be transparent between players: as such, it is related to  \cite{Borgers_1993}, \cite{Bonanno_2008}, and \cite{Bonanno_Tsakas_2018}, which we alluded to before already. In using the tools of \emph{epistemic game theory} by starting from explicitly defined assumptions concerning the players, it is related to the literature on the topic---broadly---as in \cite{Perea_2012} or \cite{Dekel_Siniscalchi_2015}  and---more precisely---to  \cite{Mariotti_2003} and \cite{Mariotti_et_al_2005}. It is related to the two latter works also in how \emph{beliefs} are coarsely represented as subset of actions (profiles). As a matter of fact, with respect to this point, it is also related to \cite{Aumann_1999a}, \cite{Samet_2010}, \cite{Bonanno_2015}, and---taking into account a different stream of literature---\cite{Chen_Micali_2015}, \cite{Chen_et_al_2015a}, \cite{Jakobsen_2020}, and \cite{Nikzad_2021}. Finally, regarding the fact that here we investigate `extreme' players' \emph{attitudes}, it is related to  \cite{Yildiz_2007} and \cite{Weinstein_2016}, where the relation with the latter arises in the way in which these attitudes are identified as polar opposites. In \Sref{sec:discussion}, we address in a more detailed way the relation between our work and some of the most closely related, aforementioned contributions. 

Tangentially related to our work, \cite{Brunner_et_al_2021} experimentally find that players more often play according to the $\max \max$ and $\max \min$ decision criteria relative to Nash equilibrium behavior.\footnote{They do not consider play according to (iterated) mutual belief of optimism and pessimism, but---within our language---their setting would correspond to the assumption that players consider all the opponents' actions as possible (which would correspond to a form of full-support assumption).} \cite{Eichberger_Kelsey_2014} study optimism and pessimism in games, but  in a setting of ambiguity and equilibrium: therefore, their contribution is distinct and complementary to our approach. Close to the paper just mentioned, \cite{Dominiak_Guerdjikova_2021}  study optimism and pessimism from a decision-theoretic standpoint by linking these notions to the study of ambiguity, whereas \cite{Schipper_2021} studies them from an evolutionary standpoint by further investigating their behavioral implications in submodular and supermodular games with aggregate externalities. Also, \cite{Guo_Yannelis_2021} study full implementation with Wald-type maxmin preferences. Finally, \cite{Gossner_Kuzmics_2019} study decision makers that ignore the actual consequences related to their choices.

\subsection{Synopsis}

Summarizing \Sref{subsec:motivation_results} in a more compact way, this paper is structured as follows. In \Sref{sec:epistemic_static}, we introduce the variety of games we study and the epistemic structures appropriate for our analysis along with our events of interest. In \Sref{sec:forms_of_rationalizability}, we define the solution concepts that algorithmically characterize the behavioral implications of the events which are the focus of our analysis. In \Sref{sec:Yildiz_2007}, we study the relation between our notion of optimism and  that of wishful thinking as introduced by \cite{Yildiz_2007}. In \Sref{sec:B-dominance}, we relate our work to the notion of B{\"o}rgers  dominance, in light of the relation between this notion and the implications of  rational behavior in ordinal games. In \Sref{sec:rationalizability}, we show how our work relates to \cite{Weinstein_2016}. Finally, in \Sref{sec:discussion}, we further discuss various aspects of our work and how our results relate to the existing literature. All the proofs of the results established in the paper are relegated to \Sref{app:proofs}.

\section{Primitive Objects} 
\label{sec:epistemic_static}

The primitive objects of our analysis are finite ordinal games. In particular, a finite \emph{ordinal game} (henceforth, game) is a tuple
\begin{equation*}
\Gamma := \la I , (A_i, \succsim_i)_{i \in I} \ra 
\end{equation*} 
where, for every $i \in I$, $A_i$ is player $i$'s finite set of \emph{actions}, with $A_{-i} := \prod_{j \in I \setminus \{i\}} A_j$ and $A := \prod_{j \in I} A_j$, and  $\succsim_i \, \subseteq A \times A$ is player $i$'s complete and transitive preference relation over action profiles. Trivially, any complete and transitive preference relation $\succsim_i$ can be represented by a utility function $u_i : A \to \Re$, which is unique up to monotone transformation. Fixing for every player one of these utility functions induces a \emph{standard game} $\Lambda(\Gamma) := \la I , (A_i, u_i)_{i \in I} \ra $. To ease notation, we slightly abuse it by using $\Gamma := \la I , (A_i, u_i)_{i \in I} \ra $ for an ordinal game, where it is understood that $u_i$ is one possible representation of $\succsim_i$. It is clear that all definitions and results about ordinal games refer to $\succsim_i$ and do not depend on the choice of the utility functions representing the preferences. However, this makes the notation slightly less involved and should not lead to any confusion. With this in mind, we call a game (ordinal or standard) \emph{generic} if $a_i \neq a'_{i}$ implies that  $u_i (a_i, a_{-i}) \neq u_i (a'_i, a_{-i})$, for every  $i \in I$, $a_i , a'_i \in A_i$, and  $a_{-i} \in A_{-i}$. Finally, for every ordinal game $\Gamma$ there exists an equivalence class of standard games induced by it, i.e., there exists an equivalence relation $\sim$ such that $\Lambda(\Gamma) \sim \Lambda'(\Gamma)$, for every two induced standard games $\Lambda(\Gamma)$ and $\Lambda'(\Gamma)$.\footnote{Clearly, we could have started with standard games as primitive objects by defining ordinal games as equivalence classes of  $\sim$. However, since our main focus is on ordinal games, we opted for using ordinal games as our primitive objects.}

In what follows, every topological space is assumed to be compact Hausdorff, where in the case of finite spaces this is a consequence of endowing them---as we do---with the discrete topology. Thus, given an arbitrary space $X$, we let $\mscr{K} (X)$ denote the family of all its nonempty, compact subsets endowed  with the Hausdorff topology,\footnote{Recall that the Hausdorff topology is the topology generated by all subsets of the form $\Set { \kappa \in \mscr{K} (X) | \kappa \subseteq G }$ and $\Set { \kappa \in \mscr{K} (X) | \kappa \cap G \neq \emptyset }$ with $G$ open in $X$.} which makes it compact Hausdorff, whenever $X$ is compact Hausdorff.\footnote{See \Sref{subsec:technical_assumptions} for a discussion of this technical assumption.}

\begin{definition}[Epistemic Possibility Structure]
\label{def:epistemic_possibility_structure}
Given a game  $\Gamma := \la I, (A_i, u_i)_{i \in I} \ra$, an \emph{epistemic possibility structure} (henceforth, possibility structure) appended to $\Gamma$ is a tuple
\begin{equation*}
\mfrak{P} := \la I , (A_{-i} , T_{i} , \pi_{i} )_{i \in I} \ra
\end{equation*}
where, for every $i \in I$, $T_i$ is her compact Hausdorff set of \emph{epistemic types} (henceforth, types) and $\pi_i : T_i \to  \mscr{K} (A_{-i} \times T_{-i})$ is her continuous \emph{possibility function}.
\end{definition}

To ease the reading, we introduce the function $\fob_i : T_i \to \mscr{K} (A_{-i})$ defined as $\fob_i (t_i) := \proj_{A_{-i}} \pi_i (t_i)$, for every $t_i \in T_i$ (where $\proj$ denotes the---continuous---projection operator  as canonically defined), which captures what an arbitrary player $i$ considers possible regarding \emph{only} the actions of the remaining players, i.e., her first-order beliefs, with the understanding that such actions could form a non-product set. For every $i \in I$, we let $\Omega_i := A_i \times T_i$, with $\Omega := \prod_{j \in I} \Omega_j$ being the \emph{state space} associated to the possibility structure and $\Omega_{-i} := \prod_{j \in I \setminus \{i\}} \Omega_j$. We call \emph{events} only the following objects, that have by construction a  product structure: $E_i \in \mscr{K} (\Omega_i)$,  $E_{-i} := \prod_{j \in I \setminus \{i\}} E_j \in \mscr{K} (\Omega_{-i})$, and $E :=  \prod_{j \in I} E_j  \in \mscr{K} (\Omega) $ (e.g., $E_i$ is an event concerning player $i$).  That is, all our events of interest are assumed to have a product structure relative to the player indices, with the understanding that $E_i$ may not be a product set across $A_i$ and $T_i$, for every $i \in I$.

\begin{remark}[Universality]\label{remark:uni}
For every game $\Gamma$, we let  $\mfrak{P}^* := \la I , (A_{-i} , T^*_{i} , \pi_{i} )_{i \in I} \ra$ denote the \emph{canonical hierarchical structure} appended to $\Gamma$ that is constructed as the space that  comprises all the players' infinite hierarchies of beliefs that satisfy a coherency requirement (see \citet[Section 3]{Mariotti_et_al_2005}). The canonical hierarchical structure $\mfrak{P}^*$ is a possibility structure in its own rights that is \emph{universal} according to the terminology introduced by \cite{Siniscalchi_2008}, that is:
\begin{itemize}[leftmargin=*]
\item it is  \emph{terminal}, since every other possibility structure can be uniquely embedded in it, 

\item and \emph{belief-complete}, since the possibility function $\pi_i$ is surjective, for every $i \in I$.
\end{itemize}
Thus, we call $\mfrak{P}^*$ the \emph{universal possibility structure}.
\end{remark} 

Given a possibility structure $\mfrak{P}$ with state space $\Omega$, interactive reasoning is captured by means of opportune modal operators acting on $\Omega$. In particular, the \emph{belief operator}\footnote{See \citet[Sections 10.2 \& 10.3]{Morris_1997} for a decision-theoretic foundation based on the notion of Savage-null events of the belief operator in the context of Aumann structures.} $\Bel_{i}$ of player $i$ is defined as
\begin{equation*}
\Bel_{i} (E_{-i}) := %
\Set { \big(a_i , t_i \big) \in A_i \times T_i | \pi_{i} (t_i) \subseteq E_{-i} } ,
\end{equation*}
for every $E_{-i} \in \mscr{K} (\Omega_{-i})$, with $\Bel (E) := \prod_{j \in I} \Bel_j (\proj_{A_{-j} \times T_{-j}} E)$ denoting the \emph{mutual belief} operator and
\begin{equation*}
\CB (E) := E \cap \Bel (E)
\end{equation*}
denoting the \emph{correct (mutual) belief} operator,  for $E \in \mscr{K} (\Omega)$.

The basic events we want to formalize concerning behavior in a possibility structure are those that capture a player being either pessimistic or optimistic. To formalize these notions, we now introduce two best-reply correspondences: in particular, given a game $\Gamma$, a player $i \in I$, and a $\kappa_i \in \mscr{K} (A_{-i})$, we let 
\begin{equation*}
\label{eq:br_opt}
\obr_i (\kappa_i) := %
\arg \max_{a_i \in A_i}  \max_{\widetilde{a}_{-i} \in \kappa_i} %
u_i (a_i , \widetilde{a}_{-i}) 
\end{equation*}
denote the set of \emph{optimistic best-replies} to belief $\kappa_i \in \mscr{K} (A_{-i})$ and we let 

\begin{equation*}
\label{eq:br_pes}
\pbr_i (\kappa_i) := %
 \arg \max_{a_i \in A_i}  \min_{\widetilde{a}_{-i} \in \kappa_i} %
u_i (a_i , \widetilde{a}_{-i}) 
\end{equation*}
denote the set of \emph{pessimistic best-replies} to belief $\kappa_i \in \mscr{K} (A_{-i})$, where we deem \emph{justifiable} all those actions belonging to $\obr_i (\kappa_i)$ or $\pbr_i (\kappa_i)$ for a given $\kappa_i \in \mscr{K} (A_{-i})$. 

Thus, we let 
\begin{equation}
	\label{eq:optimism}
	\Opt_i := \Set { (a^{*}_i, t_i) \in A_i \times T_i | %
		a^{*}_i \in \obr_i ( \fob_i (t_i)) }
\end{equation}
be the event in $\Omega_i$ that captures player $i$ being \emph{optimistic}, whereas we let 
\begin{equation}
	\label{eq:pessimism}
	\Pes_i := \Set { (a^{*}_i, t_i) \in A_i \times T_i | %
		a^{*}_i \in \pbr_i ( \fob_i (t_i)) }
\end{equation}
be the event in $\Omega_i$ that captures player $i$ being \emph{pessimistic}. 

\begin{example}[label=ex:leading_example_static, name=Leading Example]
Consider the game represented in \Sref{fig:leading_example_static} with two players, namely, Ann (viz., $a$) and Bob (viz., $b$).

\begin{figure}[H]
\centering
\begin{game}{3}{3}[$a$][$b$]
		& $L$ 	& $C$ & $R$\\
$U$ 	& $2, 3$ 	& $3, 2$ & $1, 1$\\
$M$ 	& $4, 3$ 	& $1, 1$ & $4, 0$\\
$D$ 	& $2, 0$ 	& $2, 2$ & $1, 1$\\
\end{game}
\caption{A $3\times3$ game.}
\label{fig:leading_example_static}
\end{figure}

\noindent To see the events we have introduced at work, we append to it a possibility structure. In particular, we focus on Ann, with $T_a := \{ t_a, t'_a , t''_a \}$ and
\begin{align*}
&\fob_a (t_a) := \{ L\}, \\
&\fob_a (t'_a) := \{ C \}, \\
&\fob_a (t''_a) := A_b.
\end{align*}
Then it is straightforward to observe that 
\begin{align*}
& \Opt_a := \{ (M, t_a) , (U, t'_a), (M, t''_a)\}, \\
& \Pes_a := \{ (M, t_a) , (U, t'_a), (U, t''_a) , (M, t''_a) ,(D, t''_a) \}.
\end{align*}
Crucially, the difference between Ann's attitude arises when she contemplates the idea that Bob can play more than one action, i.e., when her type is $t''_a$. If she is optimistic, she is going to expect Bob to play $L$ or $R$, because both those actions can give her the highest utility, thus she is going to play $M$ (indeed, in both cases she can get $4$); if she is pessimistic, she is indifferent between $U$, $M$, and $D$, since  $1$ is the lowest possible payoff she could get given $L$, $C$, or $R$.
\end{example}

Having defined what it means for a player to be either  optimistic or pessimistic by opportune events in $\Omega_i$, the natural next step is to investigate the implications of having players involved in a game interactively reason about each others. First, we have that $\Opt:=\prod_{i \in I} \Opt_i$ and $\Pes := \prod_{i \in I} \Pes_i$. Given this, we let $\CB^0 (\Opt):= \Opt$ and $\CB^0 (\Pes) :=\Pes$ denote the events that all players are optimistic and pessimistic, respectively. Concerning interactive reasoning, we then define inductively for every $m \in \bN$ with\footnote{Since there is no general consensus on the definition of $\bN$, to avoid any ambiguity, we set $\bN := \{0, 1, \dots \}$.} $m >0$ the corresponding (correct) $m^\text{th}$-order mutual belief events:\footnote{\label{footnote:introspection}Note that, as usual in models without introspective beliefs, we do impose correct beliefs to restrict behavior in addition to restricting (higher-order) beliefs. Even though they consider a standard Bayesian framework, the arguments made in the first paragraph in \citet[Section 12.3.2]{Dekel_Siniscalchi_2015} apply verbatim to our framework. This is not to be confused with imposing the Truth Axiom, which would impose correct beliefs for \emph{all} possible events. Such an additional assumption would change our results, as discussed in more detail in \Sref{sec:Yildiz_2007}. With this in mind, in what follows we do not employ the word ``correct'' unless explicitly needed.}
\begin{align*}
	\CB^m (\Opt) & := \Opt \cap  \Bel ( \CB^{m-1} (\Opt)),\\
	\CB^m (\Pes) & := \Pes \cap  \Bel ( \CB^{m-1} (\Pes) ).
\end{align*}
The role in the rest of the analysis of the events concerning common belief is such that they deserve their own definition.

\begin{definition}[Optimism/Pessimism and Common Belief in Optimism/Pessimism]
\label{def:ECBE}
Given a game $\Gamma$ and a possibility structure $\mfrak{P}$ with state space $\Omega$, the epistemic condition \emph{Optimism and Common Belief in Optimism}  is captured by the event 
\begin{equation*}
\OCBO := \CB^{\infty} (\Opt ) :=  \bigcap_{m \geq 0} \CB^m (\Opt),
\end{equation*}
while
\begin{equation*}
\PCBP := \CB^{\infty} (\Pes) := \bigcap_{m \geq 0} \CB^m (\Pes)
\end{equation*}
is the event that captures the condition \emph{Pessimism and Common Belief in Pessimism}.
\end{definition}

It is important at this stage to emphasize that the correct belief operator satisfies the following properties:
\begin{itemize}[leftmargin=*]
\item \emph{Conjunction Property:} $\CB (E \cap F) = \CB (E) \cap \CB (F)$, for every $E, F \in \mscr{K} (\Omega)$;

\item  \emph{Monotonicity Property:} if $E \subseteq F$, then $\CB (E) \subseteq \CB (F)$, for every $E, F \in \mscr{K} (\Omega)$.\footnote{The Conjunction and the Monotonicity property are satisfied by the belief and the mutual belief operator as well.}
\end{itemize}
As a result, it is immediate that $\CB^n (E) \subseteq \CB^m (E)$, for every $m, n \in \bN$ with $n > m$ and for every $E \in \mscr{K} (\Omega)$.\footnote{As emphasized in \citet[Section 12.3.1, p.633]{Dekel_Siniscalchi_2015} (with the understanding that---as in \Sref{footnote:introspection}---even if they consider a standard Bayesian framework, their arguments apply verbatim to our framework), it is due to the conjunction and the monotonicity properties that, by considering a repeated application of the mutual belief operator $\Bel$ according to the---standard---rules spelled out above, focusing---without loss of generality---on the event $\Opt$, we would have that $\CB^n (\Opt) = \bigcap_{k = 0}^n \Bel^k (\Opt) = \Opt \cap \bigcap_{k=1} \Bel^{k} (\Opt)$, with $n \in \bN$, and $\OCBO = \CB^\infty (\Opt) = \bigcap_{n \geq 0} \Bel^n (\Opt) = \Opt \cap \bigcap_{n \geq 1} \Bel^{n} (\Opt)$ (see \citet[Section 12.7.4.3, p.679]{Dekel_Siniscalchi_2015} for operators lacking these properties)). We would like to thank an anonymous referee for having emphasized the need to make this point explicit.}

Having established our events of interest, a crucial step whenever involved in an epistemic analysis is to establish that those events are actually epistemic `events' for the players. That is, we just defined $\OCBO$ and $\PCBP$, but are those events part of the language of the players? This is a crucial problem, since we want our players to reason about these very events. This is exactly what we achieve next.

\begin{proposition}
\label{prop:measurability_static}
Given a possibility structure $\mfrak{P}$ with state space $\Omega$ appended to a game $\Gamma$:
\begin{itemize}
\item[i)] for every $n \in \bN$, $\CB^n (\Opt) \in \mscr{K} (\Omega)$ and $\CB^n (\Pes) \in \mscr{K} (\Omega)$;

\item[ii)] $\OCBO \in \mscr{K} (\Omega)$ and $\PCBP \in \mscr{K} (\Omega)$.
\end{itemize}
\end{proposition}

The reason why \Mref{prop:measurability_static} is enough to establish this point is that, rather informally, given our topological assumptions, these  results amount to stating that the relevant sets are events in the \emph{measurable} sense of the term.\footnote{See \Sref{subsec:proofs_static} for a formalization of this point along with the proof of the result.}

\section{Capturing Optimism \& Pessimism} 
\label{sec:forms_of_rationalizability}

Having formalized the epistemic framework that we append to an ordinal game, it is natural to ask ourselves if we can algorithmically characterize the behavioral implications of the epistemic events of interest, with a particular attention to those defined in \Mref{def:ECBE}. The following two subsections  provide such characterization.

\subsection{The Optimistic Player}
\label{subsec:optimistic_static}

Building on the notion of optimistic best-replies, we now define a  solution concept which is essentially a formulation based on our  language of Point Rationalizability, as introduced in \citet[Section 3(b)]{Bernheim_1984}.

\begin{definition}[Point Rationalizability]
\label{def:point_r}
Fix a game $\Gamma := \la I, (A_i, u_i)_{i \in I}\ra$ and consider the following procedure, for every $i \in I$ and $m \in \bN$:
\begin{itemize}[leftmargin=*]
\item  (Step $m = 0$) $\PR^{0}_{i}  := A_i$;
\item (Step $m > 0$) Assume that $\PR^{m-1} := \PR^{m-1}_{i} \times \PR^{m-1}_{-i}$ has been defined and let 
\begin{equation}
\label{eq:p_def}
\PR^{m}_{i}  := %
\Set { a^{*}_i \in \PR^{m-1}_{i} | %
\exists  a^{*}_{-i} \in \PR^{m-1}_{-i} :   a^{*}_i \in \obr_i (\{ a^{*}_{-i} \})
}.
\end{equation}
\end{itemize}
Thus, for every $m \in \bN$, we let $\PR^{m}_{i}$ denote the set of actions of player $i$ that survive the $m$-th iteration of the Point Rationalizability procedure. Finally, 
\begin{equation*}
\PR^{\infty}_{i}  := \bigcap_{m \geq 0} \PR^{m}_{i}
\end{equation*}
is the set of actions of player $i$ that survive the Point Rationalizability procedure, with $\PR^{\infty} := \prod_{j \in I} \PR^{\infty}_{j}$ denoting the \emph{set of point rationalizable action profiles}. 
\end{definition}

Before seeing Point Rationalizability at work, it is important to recall that 
its nonemptiness has been established in \citet[Proposition 3.1]{Bernheim_1984}.\footnote{We provide this reference with the understanding that in our setting of finite games proving nonemptiness is actually trivial, whereas \citet{Bernheim_1984}  considers the more general class of compact-continuous games and establishes the corresponding (non-trivial) result.} Thus, we now go back to our leading example to see what are the behavioral predictions we obtain there via Point Rationalizability.

\begin{example}[continues=ex:leading_example_static, name=Leading Example]
To see \Mref{def:point_r} at work, we consider the game in \Sref{fig:leading_example_static}. There we have that $\PR^{1}_{a} = \{ U, M \}$ and $\PR^{1}_{b} = \{ L, C \}$ and then $\PR^{2}_{a} = \PR^{1}_{a}$ and $\PR^{2}_{b} = \{L\}$. As a result, $\PR^{3}_{a} = \{M \} = \PR^{\infty}_{a}$ and $\PR^{2}_{b} = \{L\} = \PR^{\infty}_{b}$.
\end{example}

We can now tackle the problem of the algorithmic characterization of the behavioral implications of Optimism and Common Belief in Optimism. As a matter of fact, the result that we state next settles the issue.

\begin{theorem}[Foundation of Point Rationalizability]
\label{th:tropical_p}
Fix a game $\Gamma$.
\begin{itemize}
\item[i)] If $\mfrak{P}$ is an arbitrary possibility structure appended to it, then 
\begin{equation}
\label{eq:p_m}
\proj_{A} \CB^n (\Opt) \subseteq \PR^{n+1},
\end{equation}
for every $n \in \bN$, and 
\begin{equation}
\label{eq:p_infty}
\proj_{A} \OCBO \subseteq \PR^{\infty}.
\end{equation}
\item[ii)] Given the universal possibility structure $\mfrak{P}^*$,
\begin{equation}
\label{eq:p_bc_m}
\proj_{A} \CB^n (\Opt) = \PR^{m+1},
\end{equation}
for every $n \in \bN$, and 
\begin{equation}
\label{eq:p_bc_infty}
\proj_{A} \OCBO = \PR^{\infty}.
\end{equation}
\end{itemize}
\end{theorem}

The proof is by induction, but intuitively part (i) holds because an optimistic best-reply to a belief $\kappa_i$ is also a (point) best-reply to (one of) the $i$-favorite co-players' action profiles in $\kappa_i$, whereas part (ii), conversely, is a consequence of both observing that deterministic (i.e., singleton) beliefs are just a particular form of belief in our framework\footnote{In \Sref{subsec:OR} we elaborate on the choice-equivalence of optimistic and point best-replies.} and the belief-completeness of the universal possibility structure considered. 

From the nonemptiness of Point Rationalizability and \Mref{eq:p_bc_infty},  it follows that, when we work with the universal possibility structure $\mfrak{P}^*$ by focusing on Optimism and Common Belief in Optimism, we always have nonempty behavioral predictions. We now show that this is not necessarily the case when we work with possibility structures that are not the universal one.

\begin{example}[continues=ex:leading_example_static, name=Leading Example]
We consider the game in \Sref{fig:leading_example_static} to which we append the possibility structure $\mfrak{P}$ where $T_i := \{t_i\}$ and $\pi_i (t_i) := A_{-i} \times T_{-i}$ for $i \in \{a, b\}$. Here, we have that $\Opt_a = \{ M\} \times \{t_a\}$ and $\Opt_b = \{L \} \times \{t_b\}$. Thus, since $\pi_a (t_a) = \{L, C, R\} \times \{t_b\}$, it is immediate to observe that $\pi_a (t_a) \nsubseteq \Opt_b$, i.e., $\Bel_a (\Opt_b) = \emptyset$. As a result, in this possibility structure we have that $\OCBO = \emptyset$.
\end{example}

\subsection{The Pessimistic Player}
\label{subsec:pessimistic_static}

We now introduce our algorithmic procedure that capture interactive pessimism in static games, that we call Wald Rationalizability in honor of Abraham Wald's celebrated decision criterion introduced in \cite{Wald_1950}.

\begin{definition}[Wald Rationalizability]
\label{def:vn_r}
Fix a game $\Gamma := \la I, (A_i, u_i)_{i \in I}\ra$ and consider the following procedure, for every $i \in I$ and $m \in \bN$:
\begin{itemize}[leftmargin=*]
\item  (Step $m = 0$) $\WR^{0}_{i}  := A_i$;
\item (Step $m > 0$) Assume that $\WR^{m-1} := \WR^{m-1}_i \times \WR^{m-1}_{-i}$ has been defined and let 
\begin{equation}
\label{eq:v_def}
\WR^{m}_{i}  := %
\Set { a^{*}_i \in \WR^{m-1}_{i} | %
\exists \kappa_i  \subseteq \WR^{m-1}_{-i} : %
 a^{*}_i \in \pbr_i (\kappa_i)
}.%
\end{equation}
\end{itemize}
Thus, for every $m \in \bN$, we let $\WR^{m}_{i}$ denote the set of actions of player $i$ that survive the $m$-th iteration of the Wald Rationalizability procedure. Finally, 
\begin{equation*}
\WR^{\infty}_{i}  := \bigcap_{m \geq 0} \WR^{m}_{i}
\end{equation*}
is the set of actions of player $i$ that survive the Wald Rationalizability procedure, with $\WR^{\infty} := \prod_{j \in I} \WR^{\infty}_{j}$ denoting the \emph{set of Wald rationalizable action profiles}.
\end{definition}

Regarding \Mref{eq:v_def}, it should be pointed out that requiring $\kappa_i \subseteq \WR^{m-1}_{-i}$ instead of $\kappa_i = \WR^{m-1}_{-i}$, for every $m>0$, avoids a kind of inclusion/exclusion problem in the spirit of \citet[Section 1]{Samuelson_1992} and it is in line with the idea that $\kappa_i$ is a \emph{subjective} belief, which---as such---may exclude objects that are not (yet) excluded by strategic reasoning.\footnote{We are grateful to the anonymous associate editor for having raised this point.}

Mirroring the structure of \Sref{subsec:optimistic_static}, we now state a crucial property of Wald Rationalizability (implied by \Mref{prop:P-V} below).

\begin{remark}[Nonemptiness]
\label{rem:nonemptiness_vn}
For every game $\Gamma$, $\WR^{\infty} \neq \emptyset$.
\end{remark}

Again, we go back to our leading example to see how Wald Rationalizability performs there.

\begin{example}[continues=ex:leading_example_static, name=Leading Example]
To see \Mref{def:point_r} at work, we consider again the game in  \Sref{fig:leading_example_static}. Now using \Mref{eq:v_def} gives $\WR^{1}_{a} = A_a$ and $\WR^{1}_{b} = \{L, C\}$. As a matter of fact, the algorithm stops here. Thus, we have that $\WR^{\infty}_{a} = A_a$ and $\WR^{\infty}_{b} = \{L, C\}$.
\end{example}

As we did for Optimism and Common Belief in Optimism, we now solve the issue of providing an algorithmic characterization for the behavioral implications of Pessimism and Common Belief in Pessimism.

\begin{theorem}[Foundation of Wald Rationalizability]
\label{th:tropical_vn}
Fix a game $\Gamma$.
\begin{itemize}
\item[i)] If $\mfrak{P}$ is an arbitrary possibility structure appended to it, then 
\begin{equation}
\label{eq:v_m}
\proj_{A} \CB^n (\Pes) \subseteq \WR^{n+1},
\end{equation}
for every $n \in \bN$, and 
\begin{equation}
\label{eq:v_infty}
\proj_{A} \PCBP \subseteq \WR^{\infty}.
\end{equation}
\item[ii)] Given the universal possibility structure $\mfrak{P}^*$,
\begin{equation}
\label{eq:v_bc_m}
\proj_{A} \CB^n (\Pes) = \WR^{n+1},
\end{equation}
for every $n \in \bN$, and 
\begin{equation}
\label{eq:v_bc_infty}
\proj_{A} \PCBP = \WR^{\infty} .
\end{equation}
\end{itemize}
\end{theorem}

As it is for the case of Point Rationalizability and Optimism and Common Belief in Optimism addressed in \Sref{subsec:optimistic_static}, it follows from the nonemptiness of Wald Rationalizability and \Mref{eq:v_bc_infty} that, when we work with the universal possibility structure $\mfrak{P}^*$ and the focus is on Pessimism and Common Belief in Pessimism, we always have nonempty behavioral predictions. We now show that this is not necessarily the case when we work with possibility structures that are not the universal one.

\begin{example}[continues=ex:leading_example_static, name=Leading Example]
We consider the game in \Sref{fig:leading_example_static} to which---once more---we append the possibility structure $\mfrak{P}$ where $T_i := \{t_i\}$ and $\pi_i (t_i) := A_{-i} \times T_{-i}$ for $i \in \{a, b\}$. Here, we have that $\Pes_a = \{ U, M, D\} \times \{t_a\}$ and $\Pes_b = \{C \} \times \{t_b\}$. Thus, since $\pi_a (t_a) = \{L, C, R\} \times \{t_b\}$, it is immediate to observe that $\pi_a (t_a) \nsubseteq \Pes_b$, i.e., $\Bel_a (\Pes_b) = \emptyset$. As a result, in this possibility structure we have that $\PCBP = \emptyset$.
\end{example}

\subsection{Relation between the Algorithms}
\label{subsec:relation_static}

Having formalized procedures that, as shown, algorithmically characterize the behavior corresponding to the epistemic events  of interests, it is natural to investigate what is the relation between the two solutions concepts just introduced. Our \Mref{ex:leading_example_static} already shows  that $\WR^\infty \not\subseteq \PR^{\infty}$. But what about the reverse inclusion? Can we say that $\PR^\infty$ is a refinement of $\WR^\infty$? On intuitive grounds, this should be the case and the following result formally establishes exactly this point.

\begin{proposition}
\label{prop:P-V}
Given a game $\Gamma$, $\PR^n \subseteq \WR^n$, for every $n \in \bN$.
\end{proposition}

The proof is equally intuitive: if a strategy is an optimistic best-reply, then it is a point best-reply to the player's favorite opponent's action as already mentioned above,\footnote{Again, we discuss this (and consequences thereof) in more detail in \Sref{subsec:OR}.} but then it also a pessimistic best-reply to the singleton belief considering only this opponent's strategy as possible. In other words,  for singleton beliefs the two notions coincide and for the optimistic case it is without loss to consider such singleton beliefs.\footnote{Besides the discussion in \Sref{subsec:OR}, we further exploit this observation in \Sref{subsec:Mariotti_2003} to shed light on the connections to \cite{Mariotti_2003}.} Conversely, a pessimistic best-reply might need a non-singleton belief. Therefore, there are occasions in which the inclusion is strict.

\section{Wishful Thinking Revisited}
\label{sec:Yildiz_2007}

\cite{Yildiz_2007} proposes a model of wishful thinking in strategic environments to which our notion of optimism shares  its behavioral attitude along with its mathematical formalization as in \Mref{eq:optimism}. However, there are some crucial differences between our approach and that of \cite{Yildiz_2007}. Most obviously, the  algorithm in \citet[Section 3]{Yildiz_2007} differs from Point Rationalizability, since the former deletes \emph{actions profiles}, while the latter \emph{actions}.

\begin{definition}[Wishful Thinking Procedure]
\label{def:ywt_p}
Fix a game $\Gamma := \la I, (A_i, u_i)_{i \in I}\ra$ and consider the following procedure, for every $m \in \bN$:
\begin{itemize}[leftmargin=*]
\item  (Step $m = 0$) $\YR^{0}  := A$;
\item (Step $m > 0$) Assume that $\YR^{m-1}$ has been defined and let 
\begin{equation*}
\label{eq:y_def}
\YR^{m}  := %
\Set { a^{*} \in \YR^{m-1} | %
\begin{array}{l}
\forall i \in I  \ 
\exists a_{-i} \in A_{-i} : \\%
1.\ (a_i^*, a_{-i}) \in  \YR^{m-1}, \\
2.\  a^{*}_i \in \obr_i (\{a_{-i}\}), \\
\displaystyle 3.\  u_i (a^{*}_i, a_{-i}) \geq \max_{a_i \in A_i} u_i (a_i, a^{*}_{-i})
\end{array}
}.
\end{equation*}
\end{itemize}
Thus, for every $m \in \bN$, $\YR^{m}$ denotes the set of action profiles that survive the $m$-th iteration of the Wishful Thinking procedure. Finally, 
\begin{equation*}
\YR^{\infty}  := \bigcap_{m \geq 0} \YR^{m}
\end{equation*}
is the \emph{set of Wishful Thinking action profiles}.
\end{definition}

\citet{Yildiz_2007} defined the algorithm in terms of deleting action profiles on each round. In contrast, our \Mref{def:ywt_p} is defined as collecting actions that are justifiable by means of wishful thinking taking as given those profiles that are deemed justifiable in the previous rounds. Of course, the difference is just a change in quantifiers, but we opted for the current version to facilitate the comparison with the procedures introduced before. In particular, the current definition makes clear the connection to Point Rationalizability (as in \Mref{def:point_r}). Indeed, the Wishful Thinking procedure is a refinement of Point Rationalizability.

\begin{proposition}
	\label{prop:Y-P}
	Given a game $\Gamma$, $\YR^{n} \subseteq \PR^n$, for every $n \in \bN$.
\end{proposition}

Furthermore, \citet[Example 5, pp.334--335]{Yildiz_2007} illustrates an existence failure of his model, whereas Point Rationalizability is always nonempty. Thus, the inclusion in \Mref{prop:Y-P} might be strict and, as a result, the behavioral implications of Optimism and Common Belief in Optimism differ from those obtained via the Wishful Thinking procedure. For illustration purposes, we now show an example of the Wishful Thinking procedure selecting a strict subset of action profiles of those selected via Point Rationalizability.

\begin{example}[label=ex:leading_example_yildiz, name=Battle of the Sexes]
To see the difference, consider the leading example of \citet{Yildiz_2007}, which happens to be the Battle of the Sexes.

\begin{figure}[H]
		\centering
		\begin{game}{2}{2}[$a$][$b$]
			& $L$ 	& $R$ \\
			$U$ 	& $2, 1$ 	& $0, 0$\\
			$D$ 	& $0, 0$ 	& $1, 2$\\
		\end{game}
		\caption{Battle of the Sexes.}
		\label{fig:leading_example_yildiz}
\end{figure}

\noindent Clearly, $\PR^\infty = A_a \times A_b$. However, the algorithm in \cite{Yildiz_2007}  deletes the action profile $(D, L)$.\footnote{Somewhat betraying the spirit of this section by focusing on a non-rich possibility structure that \emph{a fortiori} does not give $\PR^{\infty}$ as its behavioral predictions, it should be observed that this very example allows us to show that $\proj_{A} \OCBO \subsetneq \PR^{\infty}$ in the possibility structure $T_i := \{t_i\}$, for $i \in \{a, b\}$, with $\pi_a (t_a) = \{ (L, t_b)\}$ and $\pi_b (t_b) = \{ (U, t_a)\}$. As a result, in this case we would have that $\proj_{A} \OCBO \neq \PR^{\infty} \neq \YR^\infty$. We are thankful to an anonymous referee for having raised the issue of the possibility of having different behavioral predictions via $\proj_{A} \OCBO$, $\PR^\infty$, and $\YR^\infty$.}  To justify the profile $(D,L)$ under optimism/wishful thinking, Ann must believe that $L$ is \emph{impossible}, which \emph{ipso facto} is a \emph{wrong} belief, which is not allowed in the model of \citet{Yildiz_2007}, but is allowed in our framework. Indeed, as pointed out in \citet[Section 1, p.321]{Yildiz_2007}, by focusing---without loss of generality---on Ann, it is not possible for her to indulge in wishful thinking, play $D$, and \emph{believe} that $L$ is possible, since then she would believe that $L$ \emph{will} happen.\footnote{An anonymous referee suggested this statement that is clearer than the statement we had in an earlier version of this paper. We are thankful for this suggestion.}  
\end{example}

Given this example and the fact that the baseline assumptions about players' behavior are essentially the same, it is natural to ask ourselves why this difference arises with respect to  behavioral predictions. As already hinted in \Mref{ex:leading_example_yildiz}, the crucial issue lies in the modal operators employed: we use the \emph{belief} operator, while \cite{Yildiz_2007} uses  the \emph{knowledge} operator. It is well known that knowledge differs from belief in that knowledge satisfies the Truth Axiom, which states that whatever is known must be true.\footnote{See for example \citet[Section 5.1.2, p.70]{Osborne_Rubinstein_1994}.} Since belief does not  satisfy this axiom, a player in our model might believe an event that is actually wrong.\footnote{\citet[Section 3.2]{Samet_2013} provides a detailed discussion of the differences within the framework of belief structures.}

We now formalize the informal argument sketched in the paragraph above. Thus, first of all, we introduce the appropriate epistemic model to work with knowledge and then proceed by providing a proper comparison between the two approaches.\footnote{To simplify notation, in this section we focus on finite epistemic models. For the purpose of a meaningful comparison between the approaches, this restriction is without loss of generality.} 

\begin{definition}[Knowledge Structure]
\label{def:knowledge_structure}
Given a game  $\Gamma := \la I, (A_i, u_i)_{i \in I} \ra$ , a \emph{knowledge  structure} appended to $\Gamma$ is a tuple
\begin{equation*}
\mfrak{K} := \la I , \Psi, (T_i, \Pi_{i} )_{i \in I} \ra ,
\end{equation*}
where 
\begin{enumerate}[leftmargin=*]
\item for every player $i \in I$, $T_i$ is a finite set of types for each player,

\item $\Psi \subseteq \prod_{i \in I} (A_i \times T_i)$ is the state space,

\item for every $i \in I$, $\Pi_i$ is a partition of $\Psi$ with $\Pi_i(\omega) \subseteq \Psi$ denoting the cell containing $\omega \in \Psi$, and

\item for every $i \in I$, $\Pi_i$ satisfies the following properties:
	\begin{enumerate}[leftmargin=*]
		\item (Introspection) for every $\omega \in \Psi$, $\proj_{A_i \times T_i} \Pi_i (\omega) =  \{\proj_{A_i \times T_i} \omega\}$, and
		\item (Independence) for every $\omega, \omega' \in \Psi$, if $\proj_{T_i} \omega = \proj_{T_i} \omega'$, then $\proj_{A_{-i} \times T_{-i}}\Pi_i (\omega) = \proj_{A_{-i} \times T_{-i}}\Pi_i (\omega')$.
	\end{enumerate}
\end{enumerate}
\end{definition}

For readers familiar with usual definitions of knowledge structures as introduced in \cite{Aumann_1976} (henceforth,  \emph{Aumann structures}), \Mref{def:knowledge_structure} might look a bit obscure and therefore some remarks are in order. In contrast to Aumann structures, in our knowledge structures, states are not completely abstract, but rather are comprised of action-type pairs of every player similarly to states in our possibility structures. Indeed, this is the main reason why we use this definition of knowledge structure, as it makes the comparison to our possibility structures more transparent. However, in contrast to our possibility structures, but exactly as in Aumann structures, \Mref{def:knowledge_structure} allows for a state space that does not have a product structure.\footnote{See \Sref{subsec:knowledge} for a discussion of the need to consider state spaces without a product structure.} Finally, in our knowledge structures, partitions are assumed to satisfy two properties: Condition $4(a)$ states that a player is introspective in the sense of knowing his own action-type pair.\footnote{See \Sref{subsec:knowledge} for an alternative formulation of Introspection. In Aumann structures, Introspection is essentially captured via measurability assumptions.} Furthermore, we impose an independence condition as stated in Condition $4(b)$: this is due to the aforementioned intrinsic meaning of states in our formalization and is related to the arguments made in \citet[Footnote 5, p.35]{Stalnaker_1998}. This independence condition itself is reminiscent of the \emph{AI condition} of \citet[Definition 12.15, p.644]{Dekel_Siniscalchi_2015}.\footnote{See \Sref{subsec:knowledge} for a discussion of the conceptual reasons behind the decision of imposing the independence condition. Regarding the AI condition, see \cite{Bach_Perea_2020} and \cite{Guarino_Tsakas_2021} for an analysis of  its implications.}

With this in mind, any knowledge structure naturally gives rise to a possibility structure by using the same type spaces for every player and by defining the possibility functions as $\pi_i(t_i) := \proj_{A_{-i} \times T_{-i}} \Pi_i (\omega)$ for every $\omega \in \Psi$ such that $\omega = (a_i, t_i, a_{-i}, t_{-i})$, where this construction is well-defined thanks to the independence condition. The state space $\Omega$ associated with the resulting possibility structure might in general be larger than the state space associated with the knowledge structure $\Psi$: In particular, if $\Psi$ has a non-product structure, then $\Psi \varsubsetneq \Omega$. Conversely, starting from a possibility structure, it might not always be possible to construct a knowledge structure. Naturally, one would try to construct partitions with cells of the form $\{(a_i,t_i)\} \times \pi_i(t_i)$, which would satisfy introspection and independence. However, it is well known that such a construction does not yield a partition unless more restrictions are placed on the possibility functions $\pi_i$.\footnote{In particular, an appropriate version of \emph{reflexive} and \emph{Euclidean} possibility functions would be needed to obtain a partition (see \citet[Section 2]{Battigalli_Bonanno_1999} for the related definitions).}

We can now introduce the operator of interest in this framework, namely, the \emph{knowledge operator} $\K_{i}$ of player $i$,  defined as
\begin{equation*}
\K_{i} (E) := %
\Set { \omega \in \Psi | \Pi_i (\omega) \subseteq E } ,
\end{equation*}
for a (possibly non-product) $E \in \mscr{K} (\Psi)$. Naturally,  $\K (E) := \cap_{i \in I} \K_i (E)$ and the iterated application of the operator gives rise to $\K^m (E)$. Hence, $\mbb{K}^{\infty} (E)$ denotes the \emph{common knowledge operator} applied on an arbitrary event $E \in \mscr{K} (\Psi)$.

It is important to observe that we do not need to define a correct knowledge operator. Indeed, the operator $\mbb{K}$ satisfies the so called \emph{Truth Axiom}, i.e., $\K_i (E) \subseteq E$, for every (possibly non-product) $E \in \mscr{K} (\Omega)$. In other words, whatever a player knows is also true. Hence, a correct knowledge operator would be redundant, since knowledge implies being correct.\footnote{In contrast, when working with possibility structures, correctness has to be imposed for some particular events to restrict behavior. \Sref{footnote:introspection} discusses this point in more detail.} This difference is critical for the dichotomy optimism/wishful thinking and illustrates the discrepancies in the behavioral implications for the Battle of the Sexes. 

\begin{example}[continues=ex:leading_example_yildiz, name=Battle of the Sexes]
Consider again the game depicted in \Sref{fig:leading_example_yildiz}. We append a possibility structure to it with $T_a:=\{t_a, t_a^\prime\}$, $T_b:=\{t_b, t_b^\prime\}$, and
	\begin{align*}
		\pi_a(t_a) &= \{(L, t_b), (R, t_b^\prime)\}, &\quad&& %
		\pi_a (t_a^\prime) &= \{(R, t_b^\prime)\},\\
		\pi_b(t_b) &= \{(U, t_a)\}, &\text{ and }&& %
		\pi_b(t_b^\prime) &= \{(U, t_a),(D, t_a^\prime)\}.
	\end{align*}
Within this possibility structure, we have $\Opt_a = \{ (U, t_a), (D, t_a^\prime) \}$ and  $\Opt_b = \{ (L, t_b), (R, t_b^\prime) \}$. Because these states are the only ones which are considered possible by the players, there is optimism and common belief in optimism. In particular, note that the behavioral implications correspond to $\PR^\infty = A_a \times A_b$. Now, let us have a close look at the state $\big((D, t_a^\prime), (L, t_b)\big)\in  \OCBO$. At this state, since $\pi_a (t'_a) = \{(R, t_b^\prime)\}$,  Ann clearly holds a  wrong belief. Therefore, Ann cannot know $\{(R, t_b^\prime) \}$ at this state as this would violate the Truth Axiom. Thus, any event she knows at this state has to be a strict superset of $\{(R, t_b^\prime)\}$ and---in particular---has to include Bob's action $L$.  Wishful thinking in \cite{Yildiz_2007} is defined with respect to knowledge. Therefore, at this state she cannot choose $D$ as a wishful thinker \emph{\`{a} la} \cite{Yildiz_2007}. This argument generalizes leading to a removal according to the algorithm in \cite{Yildiz_2007}.
\end{example}

We can now translate in our language \citet[Proposition 1]{Yildiz_2007} in terms of \emph{common knowledge of optimism}.\footnote{\label{foot:opt_Yildiz}Pedantically, we should also define a new event corresponding to optimism in presence of knowledge functions, since possibility functions enter in the definition of optimism as in \Mref{eq:optimism}. For a knowledge structure, the corresponding event would be
	\begin{equation*}
		\Set { \omega^*=(a_i^*, t_i^*, a_{_i}^*, t_{_i}^*) \in \Psi | %
			a^{*}_i \in \arg \max_{a_i \in A_i} %
			\max_{\widetilde{a}_{-i} \in \proj_{A_{-i}} \Pi_i(\omega^*)} %
			u_i (a_i, \widetilde{a}_{-i}) },
	\end{equation*}
to which we do not assign a new symbol to avoid further notational clutter.}  

\begin{theorem}
\label{th:tropical_wt}
Fix a game $\Gamma$.
\begin{itemize}
\item[i)] If $\mfrak{K}$ is an arbitrary knowledge structure appended to it, then 
\begin{equation*}
\proj_{A} \K^n (\Opt) \subseteq \YR^{n+1},
\end{equation*}
for every $n \in \bN$, and 
\begin{equation*}
\proj_{A} \K^{\infty} (\Opt) \subseteq \YR^{\infty}.
\end{equation*}
\item[ii)] There exists a knowledge structure $\mfrak{K}$ such that,
\begin{equation*}
\proj_{A} \K^n (\Opt) = \YR^{n+1},
\end{equation*}
for every $n \in \bN$, and 
\begin{equation*}
\proj_{A} \K^{\infty} (\Opt)  = \YR^{\infty} .
\end{equation*}
\end{itemize}
\end{theorem}

\section{Relation to B{\"o}rgers Dominance}
\label{sec:B-dominance}

We now compare the behavior of Point Rationalizability and Wald Rationalizability to a form of Rationalizability built upon the notion of  B{\"o}rgers Dominance, introduced in  \citet[Definition 4, p.426]{Borgers_1993}.

Given a game $\Gamma$ and a player $i \in I$,  action $a_i \in A_i$ is \emph{weakly dominated relative to $\widetilde{A}_{-i}$}  for player $i$ by action $a^{*}_i \in A_i$ if $u_i (a^{*}_i, a_{-i}) \geq u_i (a_i, a_{-i})$ for every $a_{-i} \in \widetilde{A}_{-i}$ and there exists an action $a^{*}_{-i} \in \widetilde{A}_{-i}$ such that  $u_i (a^{*}_i, a^{*}_{-i}) > u_i (a_i, a^{*}_{-i})$.\footnote{It has to be observed that---typically---it is necessary to specify also a subset $\widetilde{A}_i \subseteq A_i$ of actions of player $i$ with respect to which admissibility is defined. Since for our purposes this is not necessary, we omit it to lighten the terminology and the notation.}  Thus,  action $a^{*}_i \in \widetilde{A}_i$ is \emph{admissible relative to} $\widetilde{A}_{-i}$ if it is not weakly dominated relative to $\widetilde{A}_{-i}$ and we let $\mbf{A}_i (\widetilde{A}_{-i})$ denote the set of actions of player $i$ that are admissible. Even if for our purposes it is enough to define admissible actions, it is instructive to recall that an action $a_i \in \widetilde{A}_i$ is \emph{B{\"o}rgers dominated  with respect to $\widetilde{A}_{-i}$} if  $a_i \notin \mbf{A}_i ( \widetilde{A}^{*}_{-i})$, for \emph{every}  nonempty subset $\widetilde{A}^{*}_{-i} \subseteq \widetilde{A}_{-i}$.

Armed with this definition, we want to formalize in our language based on `coarse' beliefs a notion of Rationalizability based on this dominance notion. To achieve this result, given a game $\Gamma$, a player $i \in I$, and a belief $\kappa_i \in \mscr{K} (A_{-i})$, we let
\begin{equation}
\label{eq:bbr}
	\bbr_i (\kappa_i) := \mbf{A}_i (\kappa_i)
\end{equation}
denote the set of \emph{admissible best-replies} to belief $\kappa_i \in \mscr{K} (A_{-i})$.

Much in the same spirit of the procedures we defined in the previous sections, this is really everything we need to formalize in our language B{\"o}rgers Rationalizability, stated next.

\begin{definition}[B{\"o}rgers Rationalizability]
\label{def:IBD}
Fix a game $\Gamma := \la I, (A_i, u_i)_{i \in I}\ra$ and consider the following procedure, for every $i \in I$ and $k \in \bN$:
\begin{itemize}[leftmargin=*]
\item  (Step $m = 0$) $\BR^{0}_{i}  := A_i$;
\item (Step $m > 0$) Assume that $\BR^{m-1} := \BR^{m-1}_i \times \BR^{m-1}_{-i}$ has been defined and let 
\begin{equation}
\label{eq:b_def}
\BR^{m}_{i} := %
\Set { a^{*}_i \in \BR^{m-1}_i | %
\exists \kappa_i \subseteq \BR^{m-1}_{-i} :\ %
a^{*}_i \in \bbr_i (\kappa_i) }.\\
\end{equation}
\end{itemize}
Thus, for every $m \in \bN$, $\BR^{m}_{i}$ denotes the set of actions of player $i$ that survive the $n$-th iteration of B{\"o}rgers Rationalizability. Finally, 
\begin{equation*}
\BR^{\infty}_{i}  := \bigcap_{m \geq 0} \BR^{m}_{i}
\end{equation*}
is the set of actions of player $i$ that survive B{\"o}rgers Rationalizability, with $\BR^{\infty} := \prod_{j \in I} \BR^{\infty}_{j}$ denoting the \emph{set of action profiles surviving B{\"o}rgers Rationalizability}.
\end{definition}

It has to be observed that B{\"o}rgers undominance as defined above is clearly \emph{not} captured in \Mref{eq:bbr}, but rather in \Mref{eq:b_def}, where the necessary union across all subsets of the $\kappa_i \in \mscr{K} (A_{-i})$ under scrutiny is taken. Given this, it is well known that $\BR^{\infty}$ in nonempty.

Now we can proceed by providing the epistemic foundation to this algorithmic procedure in our epistemic framework based on possibility structures. Before doing so, we want to highlight that already our definition of the procedure is built on having admissibility as the relevant notion of individual behavior and B{\"o}rgers (un)dominance is only a behavioral manifestation of admissibility across all possible types. Thus, for any possibility structure $\mfrak{P}$ appended to a game $\Gamma$ we let\footnote{\cite{Bonanno_Tsakas_2018} use a similar notion based on introspection.}
\begin{equation}
	\label{eq:rationality}
	\Ad_i := \Set { (a^{*}_i, t_i) \in A_i \times T_i | %
		a^{*}_i \in \mbf{A}_i (\varphi_i (t_i))  }
\end{equation}
denote the event that captures those states in $\Omega_i$ where player $i$ does choose an admissible action given her beliefs (as captured via types). Observe that, in contrast to $\Opt_i$ and $\Pes_i$, the event $\Ad_i$ is \emph{not} defined as an optimal choice for a decision criterion, but rather directly based on a dominance notion. That is, whereas our notions of optimism and pessimism are based on classic decision criteria under ignorance, admissibility is fundamentally a notion of (un)dominance. 

With the event $\Ad_i$ at our disposal we need to make sure the related events are measurable. For this define $\Ad$ and $\CB^n (\Ad)$ for every $n \in \bN$ similar to the definition about optimism and pessimism. Then, all (common belief) events about admissibility are measurable.

\begin{proposition}
\label{prop:measurability_A}
Given a possibility structure $\mfrak{P}$ with state space $\Omega$ appended to a game $\Gamma$:
\begin{itemize}
\item[i)] for every $n \in \bN$, $\CB^n (\Ad) \in \mscr{K} (\Omega)$;

\item[ii)] $\ACBA \in \mscr{K} (\Omega)$.
\end{itemize}
\end{proposition}

Now, it is straightforward to proceed with an epistemic foundation of B{\"o}rgers Rationalizability, as we do next.

\begin{theorem}[Foundation of B{\"o}rgers Rationalizability]
	\label{th:tropical_a}
	Fix a game $\Gamma$.
	\begin{itemize}
		\item[i)] If $\mfrak{P}$ is an arbitrary possibility structure appended to it, then 
		\begin{equation*}
			\label{eq:a_m}
			\proj_{A} \CB^n (\Ad) \subseteq \BR^{n+1},
		\end{equation*}
		for every $n \in \bN$, and 
		\begin{equation*}
			\label{eq:a_infty}
			\proj_{A} \ACBA \subseteq \BR^{\infty}.
		\end{equation*}
		\item[ii)] Given the universal possibility structure $\mfrak{P}^*$,
		\begin{equation*}
			\label{eq:a_bc_m}
			\proj_{A} \CB^n (\Ad) = \BR^{n+1},
		\end{equation*}
		for every $n \in \bN$, and 
		\begin{equation*}
			\label{eq:a_bc_infty}
			\proj_{A} \ACBA = \BR^{\infty}.
		\end{equation*}
	\end{itemize}
\end{theorem}

Our characterization can be seen as taking the perspective of the players. Within a different framework, \citet[Theorem 1, p.5]{Bonanno_Tsakas_2018} state a seemingly similar result, but provide a different proof. The difference can be interpreted as their analysis taking the perspective of an (outside) analyst. Therefore, we see \Mref{th:tropical_a} as  complementary to \citet[Theorem 1, p.5]{Bonanno_Tsakas_2018}.\footnote{\label{foot:complete_structures}See \citet[Sections 2.2--2.4]{Friedenberg_Keisler_2021} for a thorough discussion of these two interpretations.} 

As the---well known---result that follows establishes, it is rather easy to show that there exists an immediate relation between Point Rationalizability and B{\"o}rgers Rationalizability. Like in \Mref{prop:P-V}, the argument follows from the coincidence of the two best-replies for singleton beliefs.\footnote{We provide a direct and simple proof in the appendix without reference to standard best-replies/Rationalizability. However, this result is obvious given the well known implications of point-best-replies being best-replies, which in turn are B{\"o}rgers-undominated actions. For the latter, see \Sref{sec:rationalizability}.}

\begin{proposition}
\label{prop:P_B}
Given a game $\Gamma$, $\PR^{n} \subseteq \BR^n$, for every $n \in \bN$.
\end{proposition}

However, as the two examples that follow show, it is not possible to establish an inclusion relation between B{\"o}rgers Rationalizability and Wald Rationalizability.

\begin{example}[continues=ex:leading_example_static, name={Leading Example, $\WR^{\infty} \not\subseteq \BR^{\infty}$}]
Consider again the game depicted in \Sref{fig:leading_example_static}, where the only payoffs represented are those of Ann.  It is easy to observe that $D \notin \BR^{1}_a$. Indeed, for every singleton $\{ a_b \} \in A_b$, there exists an action in $A_a$ that strictly dominates $D$ (e.g., $U$ strictly dominates $D$ with respect to $C$; also, $U$ weakly dominates $D$ with respect to $\{ L, C\}$ and  $\{C, R\}$); $M$ strictly dominates $D$ with respect to $\{ L, R\}$; finally, $U$ strictly dominates $D$ with respect to $A_b$. However, as we already observed, $\WR^{1}_a = A_a$, since $A_a = \arg \max_{a_a \in A_a} \pbr_a (\kappa_a)$ for $\kappa_a = A_b$. 
\end{example}

\begin{example}[$\BR^\infty \not\subseteq \WR^{\infty}$]
\label{ex:B_not_V}
Consider the following game, with two players, namely, Ann (viz., $a$) and Bob (viz., $b$), where only Ann's payoffs are represented.

\begin{figure}[H]
\centering
\begin{game}{3}{2}[$a$][$b$]
		& $L$ 	& $R$\\
$U$ 	& $6$ 	& $1$\\
$M$ 	& $5$ 	& $2$\\
$D$ 	& $4$ 	& $3$\\
\end{game}
\caption{A game showing that $\BR^\infty \not\subseteq \WR^{\infty}$.}
\label{fig:example_B_not_V}
\end{figure}

\noindent It is easy to observe that $\BR^{1}_a = A_a$. However, $M \notin \WR^{1}_a$. Indeed, $\pbr_a (\kappa_a) = \{U\}$ with $\kappa_a = \{ L \}$, while $\pbr_a (\kappa'_a) = \{D\}$ with $\kappa'_a = \{ R \}$ or $\kappa'_a = \{ L, R \}$.
\end{example}

However, if the game is generic, things change and B{\"o}rgers rationalizable actions result in being a superset of the Wald rationalizable ones. The reason is simple: in generic games, B{\"o}rgers dominance is the same as strict dominance by a pure action. A strategy that is strictly dominated by a pure action cannot be pessimistic best-reply either, but, as \Mref{ex:B_not_V} shows, the converse does not hold. More generally, in non-trivial (i.e. when the opponent has more than one action available) games, the number of actions satisfying the $\max\min$-criterion is bounded by the number of nonempty subsets of $A_{-i}$, whereas no such bound exists for B{\"o}rgers-undominated actions. The following proposition summarizes this discussion formally.

\begin{proposition}
\label{prop:generic_V_B}
Given a generic game $\Gamma$, $\WR^{n} \subseteq \BR^n$, for every $n \in \bN$.
\end{proposition}

\section{Relation to  Rationalizability}
\label{sec:rationalizability}

Although one of our motivations is to study interactions under ignorance, the notions of optimism and pessimism could be seen as decision criteria under extreme risk seeking and risk aversion, respectively, when players do have beliefs in form of probability measures. Among other things, \cite{Weinstein_2016} studies the predictions of the standard Rationalizability algorithm (as in \citet[Definition 54.1, Chapter 4.1]{Osborne_Rubinstein_1994}---henceforth, Rationalizability) when players' risk attitudes vary.\footnote{\cite{Battigalli_et_al_2016} extend \citet[Lemma 3, p.1048]{Pearce_1984} by allowing the presence of ambiguity. In the corresponding working paper, the authors additionally study Rationalizability with ambiguity aversion, by also discussing the relation between their endeavor, B-dominance, and the discontinuity we study here. See also \cite{Dominiak_Schipper_2019} for a study of Rationalizability in presence of capacities.} In particular, he characterizes the limits of the algorithm if risk attitudes tend to either extremes: while Rationalizability converges to Point Rationalizability in the limiting case of extreme risk seeking behavior,  Rationalizability converges to B{\"o}rgers Rationalizability in the limiting case of extreme risk aversion. Now, it has to be observed that our definitions of Point and Wald Rationalizability can be seen as the  limit points of the convergence process described above once the opponents' actions a player considers as possible are those that belong to the support of her probabilistic belief in the standard model. With this association in mind, focusing on the most interesting case,  Pessimism and Common Belief in Pessimism  can be interpreted as extreme risk aversion as commonly believed among players. Thus, to clarify why \Mref{prop:generic_V_B} is not puzzling after all, they are simply---as anticipated in \Sref{subsec:motivation_results}---a manifestation of a discontinuity.

In light of this observation, an analysis of the relation between Wald Rationalizability and Rationalizability might be of interest for applications. However, it has to be pointed out that Rationalizability crucially relies on beliefs in the usual sense of probability measures or, equivalently due to \citet[Lemma 3, p.1048]{Pearce_1984}, on strict dominance by possibly \emph{mixed} actions. Either way, such constructs are ruled out in our setting. Hence,  it is conceptually inappropriate to compare our algorithms to Rationalizability. Nonetheless, given this caveat, we proceed with this comparison in a mechanical fashion for the potential applications that could arise.  Thus, we let $\TR^\infty$ denote the set of rationalizable actions and $\TR^{1}_i$ the collection of payer $i$'s actions surviving the first iteration of the Rationalizability algorithm. Given that $\TR^1 \subseteq \BR^1$ from \citet[Proposition, p.427]{Borgers_1993} and that induction provides the inclusion for further rounds of the procedures (as noted by \citet[p.1885]{Weinstein_2016}), we have that $\TR^\infty \subseteq \BR^\infty$ and---as a result---the discussion in \Sref{sec:B-dominance} does not provide further guidance on the relationship with $\WR^\infty$ for nongeneric games. As a matter of fact, there is no relationship even for generic games, as the next two examples show.

\begin{example}[continues=ex:B_not_V, name=$\TR^\infty \not\subseteq \WR^{\infty}$]
In the generic game of \Sref{fig:example_B_not_V}, it is easy to see that $\TR^1_a=\BR^{1}_a = A_a$, but $M \notin \WR^{1}_a$ as argued before.
\end{example}

\begin{example}[$\WR^\infty \not\subseteq \TR^\infty$]
	\label{ex:V_not_R}
Consider the following game, with two players, namely, Ann (viz., $a$) and Bob (viz., $b$), where only Ann's payoffs are represented.
	
	\begin{figure}[H]
		\centering
		\begin{game}{3}{2}[$a$][$b$]
			& $L$ 	& $R$\\
			$U$ 	& $3$ 	& $0$\\
			$M$ 	& $1$ 	& $1$\\
			$D$ 	& $0$ 	& $3$\\
		\end{game}
		\caption{A generic game showing that $\WR^\infty \not\subseteq \TR^{\infty}$.}
		\label{fig:example_V_not_R}
	\end{figure}
	
\noindent Here, $M$ is the only strategy of Ann which is strictly dominated (by a mixture of $U$ and $D$). Hence, $M \notin \TR_a^1$.  However, $M \in \WR^{1}_a$, because $M \in \pbr_a (\kappa_a)$ with $\kappa_a = \{ L,R \}$.
\end{example}

\Mref{ex:B_not_V} might suggest a failure of upper hemicontinuity of the Rationalizability correspondence taking the limit to extreme risk aversion. To appreciate this point, we recall a definition from \citet[Section 2]{Weinstein_2016}, stated next, where, as usual, $\Delta (A)$ denotes the set of all (correlated) mixed \emph{action profiles} and $\supp \mu$ denotes the support of an arbitrary probability measure $\mu \in \Delta (A)$, i.e., the set of all $a \in A$ such that $\mu [a] > 0$.

\begin{definition}
\label{def:concave_family}
Fix an (ordinal) game $\Gamma=\la I, (A_i, u_i)_{i \in I}\ra$. An indexed family of induced standard games $\Lambda^{-r}(\Gamma):= \la I, (A_i, u^{-r}_i)_{i \in I}\ra$, with $r \in (0, \infty)$, is \emph{unboundedly concave} if 
\begin{enumerate}[leftmargin=*]
\item for every $r > s$ and $i \in I$, $u^{-r}_i = f_{i, r, s} \circ u^{-s}_i$ for an increasing and concave function $f_{i, r, s}$,

\item for every $\pi , \pi' \in \Delta (A)$, if
\begin{equation*}
\min_{a \in \supp (\pi)} u_i (a) > \min_{a \in \supp (\pi')} u_i (a),
\end{equation*}
then there exists a $\widetilde{r} \in (0, \infty)$ such that $\sum_{a \in A} u^{-r}_i(a)\pi[a] > \sum_{a \in A} u^{-r}_i(a)\pi'[a]$, for every $r > \widetilde{r}$.
\end{enumerate}
\end{definition}

Also, we  define an indexed family of induced standard games $\Lambda^{r} (\Gamma)$ to be \emph{unboundedly convex} by taking  \Mref{def:concave_family} and by substituting all the instances of ``$-r$'', ``$-s$'' ``concave'', and ``$\min$'' with ``$r$'', ``$s$'', ``convex'', and ``$\max$'', respectively. In general, to simplify notation we suppress the explicit reference to the ordinal game in the indexed family, since the context should make the underlying ordinal game clear. Thus, we just write $\Lambda^{-r}$ for a generic member of such a family and when we apply Rationalizability on a member $\Lambda^{-r}$, we write $\TR^\infty ( \Lambda^{-r} )$. Similarly, we write $\BR^\infty(\Gamma)$ and $\WR^\infty(\Gamma)$ for the corresponding ordinal game to denote the action profile that are  B{\"o}rgers and Wald rationalizable, respectively.

For a given unboundedly concave family, \citet[Proposition 3, p.1888]{Weinstein_2016} proves that $\TR^{\infty} (\Lambda^{-r})$ is increasing in $r$ (by set-inclusion) and, loosely speaking,  $\lim_{r \to \infty} \TR^\infty (\Gamma^{-r}) =  %
\BR^\infty(\Gamma)$. That is, as players become more risk averse, the set of rationalizable action profiles increases and in the limit the set approaches B{\"o}rgers Rationalizability action profiles. However, \citet[p.1886]{Weinstein_2016} also observes that the limiting game itself corresponds to a game with preferences given by the $\max \min$ criterion. Thus, if the limit is taken before Rationalizability is applied to the game, we could expect Wald Rationalizability to be the appropriate solution concept, because, after all, the limiting game is one in which players have Pessimism and Common Belief in Pessimism. Equivalently, but staying informal, one would expect $\WR^\infty(\Gamma) = %
\TR^{\infty} \Big( \lim_{r \to \infty} \Lambda^{-r} \Big)$.
Now, \Mref{ex:B_not_V} illustrates that $\WR^\infty(\Gamma) \subsetneq \BR^\infty(\Gamma)$ or, in this informal language, that $\TR^{\infty} \Big( \lim_{r \to \infty} \Lambda^{-r} \Big) \subsetneq %
\lim_{r \to \infty} \TR^\infty(\Gamma^{-r})$, which---seemingly---corresponds to a failure of upper hemicontinuity mentioned above. In particular, along the sequence, $M$ is always rationalizable, but $M \notin \WR^{1}_a$. 

However, there is a problem with this informal argument. Indeed, \Mref{ex:B_not_V} does not show a failure of upper hemicontinuity, because
$\lim_{r \to \infty}\Lambda^{-r}$ might be ill-defined. As already pointed out by \citet[p.1892]{Weinstein_2016},\footnote{Observe that, although his argument is made for Nash equilibrium, it applies to the Rationalizability correspondence as well.} we might have unbounded payoffs and, therefore, the sequence of games might not have a convergent subsequence: this is exactly what happens in \Mref{ex:B_not_V}. To remedy this problem, it suffices to additionally impose \emph{normalized} payoffs in $\Lambda^{-r}$, for every $r \in (0, \infty)$: e.g., $\min_a u^{-r}_i(a)=0$ and $\max_a u^{-r}_i(a)=1$, for every $i \in I$. With this normalization, the limiting game $\Lambda^{-\infty}:=\lim_{r \to \infty}\Lambda^{-r}$ is well-defined. In what immediately follows, we use again \Mref{ex:B_not_V} to show this point.

\begin{example}[continues=ex:B_not_V, name=Limiting Game]
	Starting from \Sref{fig:example_B_not_V}, \Sref{fig:example_B_not_V_cont} shows the corresponding limiting game $\Lambda^{-\infty}$ for any unboundedly concave family with payoffs normalized to lie within $[0,1]$. 
	\begin{figure}[H]
		\centering
		\begin{game}{3}{2}[$a$][$b$]
			& $L$ 	& $R$\\
			$U$ 	& $1$ 	& $0$\\
			$M$ 	& $1$ 	& $1$\\
			$D$ 	& $1$ 	& $1$\\
		\end{game}
		\caption{Limiting game of extreme risk aversion of \Sref{fig:example_B_not_V}.}
		\label{fig:example_B_not_V_cont}
	\end{figure}
\noindent Clearly, we have that $\WR_1^a=A_a$, thus, restoring upper hemicontinuity.
\end{example}

Given the above, we can show that Rationalizability fails lower hemicontinuity, i.e., we can find a game such that even with this normalization in place we have
\begin{equation*}
	\WR^\infty(\Lambda^{-\infty}) := %
	\TR^{\infty} \Big( \lim_{r \to \infty} \Lambda^{-r} \Big) \supsetneq %
	\lim_{r \to \infty} \TR^\infty (\Gamma^{-r}) = %
	\BR^\infty(\Gamma).
\end{equation*}

\begin{example}[continues=ex:leading_example_static, name=Limiting Game]
	Consider the limiting game $\Lambda^{-\infty}$ corresponding to the game in \Sref{fig:leading_example_static}. Focusing on Bob, \Sref{fig:leading_example_cont} depicts his payoffs in this limiting game.
	
	\begin{figure}[H]
		\centering
		\begin{game}{3}{3}[$a$][$b$]
			& $L$ 	& $C$ & $R$\\
			$U$ 	& $1$ 	& $1$ & $1$\\
			$M$ 	& $1$ 	& $1$ & $0$\\
			$D$ 	& $0$ 	& $1$ & $1$\\
		\end{game}
		\caption{Limiting game of extreme risk aversion of \Sref{fig:leading_example_static}.}
		\label{fig:leading_example_cont}
	\end{figure}
	
	\noindent In this limiting game, we have $\WR^1_b=A_b$ and, in particular, $R \in \WR_b^1$. However, along the sequence $R$ will be always strictly dominated by $C$ and therefore $R$ cannot be an element of the limit of the upper-hemicontinuous Rationalizability correspondence.\footnote{Equivalently, $R$ being strictly dominated by the pure action $C$ implies $R \notin \BR^1_b$.}
\end{example}

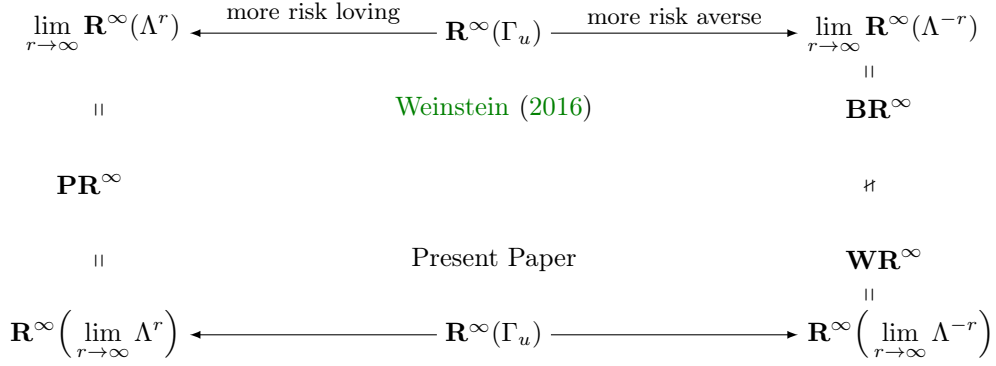
\begin{figure}[H]
\centering
\begin{tikzpicture}
		\node at (0,0) {$\TR^\infty (\Gamma_u)$};
\draw[-latex] (-0.75, 0) -- node[above] {\small more risk loving}(-4,0) node[left] {$\displaystyle \lim_{r \to \infty} \TR^\infty (\Lambda^{r})$};
\draw[-latex] (0.75, 0) -- node[above] {\small more risk averse}(4,0) node[right] {$\displaystyle \lim_{r \to \infty} \TR^\infty (\Lambda^{-r})$};
		\node at (0,-1) {\cite{Weinstein_2016}};
		\node[left] at (-4.75,-2) {$\PR^\infty$};
		\node[left] at (-5,-1) {$\shortparallel$};
		\node at (0,-4) {$\TR^\infty (\Gamma_u)$};
\draw[-latex] (-0.75,-4) -- (-4,-4) node[left] {$\displaystyle \TR^\infty \Big( \lim_{r \to \infty} \Lambda^r \Big)$};
\draw[-latex] (0.75,-4) -- (4,-4) node[right] {$\displaystyle \TR^\infty \Big( \lim_{r \to \infty} \Lambda^{-r} \Big)$};
		\node at (0,-3) {Present Paper};
		\node[left] at (-5,-3) {$\shortparallel$};
		\node[right] at (4.5,-3) {$\WR^\infty$};
		\node[left] at (5.13,-3.45) {$\shortparallel$};
		\node[right] at (4.5,-1) {$\BR^\infty$};
		\node[left] at (5.13,-0.5) {$\shortparallel$};
		\node[right] at (4.75,-2) {$\nshortparallel$};
\end{tikzpicture}
\caption{Schematic visualization of the relation between the present paper and \cite{Weinstein_2016}.}
\label{fig:relation_Weinstein}
\end{figure}

Note that in both examples, the limiting game $\Lambda^{-\infty}$ is \emph{not} induced from the ordinal game $\Gamma$ we started from, i.e., $\Lambda^{-\infty} \not\sim \Lambda^{-r}(\Gamma)$, for every $r \in (0,\infty)$. Thus, it would be inappropriate to set $\WR^\infty(\Gamma) := \TR^{\infty} \Big( \lim_{r \to \infty} \Lambda^{-r}(\Gamma) \Big)$. As a result, our analysis of $\PCBP$ along with the introduction of Wald Rationalizability clarifies the conceptual underpinnings behind the discontinuity hinted in \citet[p.1891]{Weinstein_2016} and formally illustrated in the previous example.\footnote{It should also be highlighted that \citet[p. 1891]{Weinstein_2016} rightly mentions that games with $\max\min$ preferences might ``admit no [Nash] equilibrium''. For such cases, he suggests to use the limit of the Nash equilibrium correspondence as a candidate for equilibrium in these limiting games. This issue of non-existence does not arise in our setting, because $\WR^\infty$ is always nonempty, as pointed out in \Mref{rem:nonemptiness_vn}.} In particular, \Sref{fig:relation_Weinstein} provides an immediate representation of the relation between the results established in \cite{Weinstein_2016} and those presented in the present paper.

Thus, whether one takes $\WR^\infty$ or $\BR^\infty$ as the appropriate solution concept depends on the application. If the interactive situation is best captured by $\PCBP$, then our analysis shows that $\WR^\infty(\Gamma)$ is the right solution concept. When the question is what are the behavioral implications of (common belief in) extreme risk aversion, then $\WR^\infty(\Lambda^{-\infty})$ should be used. Finally, if the analyst wants to study the limiting behavior of extreme risk aversion in a situation of (common belief) of (Bayesian) rationality,\footnote{Recall that, in this case, Rationalizability captures these behavioral implications, as shown by \cite{Brandenburger_Dekel_1987} and \cite{Tan_daCosta_1988}. See \cite{Friedenberg_Keisler_2021} for a more modern and thorough  discussion.} then $\BR^\infty(\Gamma)$ is the suitable solution concept as shown by \citet[Proposition 3, p.1888]{Weinstein_2016}.

\section{Discussion}
\label{sec:discussion}

\subsection{Optimistic Rationalizability}
\label{subsec:OR}

In this paper we focus from the outset on linking Point Rationalizability to Optimism and Common Belief in Optimism. However, it is important to observe it is possible to define another solution concept, call it \emph{Optimistic Rationalizability}, defined for every $i \in I$ as $\mbf{OR}^{0}_i := A_i$ and, assuming that $\mbf{OR}^{m-1} := \mbf{OR}^{m-1}_{i} \times \mbf{OR}^{m-1}_{-i}$ has been defined, 
\begin{equation*}
\mbf{OR}^{m}_{i}  := %
\Set { a^{*}_i \in \mbf{OR}^{m-1}_{i} | %
\exists  \kappa_i \subseteq \mbf{OR}^{m-1}_{-i} :   a^{*}_i \in \obr_i (\kappa_i)
},
\end{equation*}
for every $m>0$, with $\mbf{OR}^{\infty}_i$ and $\mbf{OR}^{\infty}$ as canonically defined, that has the property of being `symmetric' to Wald Rationalizability, as can be noticed by comparing the definition of $\mbf{OR}^{m}_i$ above and \Mref{eq:v_def}. Now, Optimistic Rationalizability is actually equivalent to Point Rationalizability. This can be shown inductively (with a trivial base case) for every $i \in I$ by observing that, for every $m>0$: $\PR^{m}_i \subseteq \mbf{OR}^{m}_i$ holds, because singleton beliefs are subsets of arbitrary nonempty beliefs $\kappa_i$; $\mbf{OR}^{m}_i \subseteq \PR^{m}_i$ holds, because for an arbitrary $a^{*}_i \in \mbf{OR}^{m}_i$ and corresponding $\kappa_i \subseteq \mbf{OR}^{m-1}_{-i}$ such that
\begin{equation*}
a_{i}^{*}\in \obr_i (\kappa_i) = \arg \max_{a_{i}\in A_{i}}\max_{a_{-i}\in
\kappa_{i}} u_{i} (a_{i},a_{-i})
\end{equation*}
we have, for an arbitrary $a_{-i}^{*}\in \arg \max_{a_{-i}\in \kappa_i} u_i (a_{i}^{*},a_{-i}) \subseteq \kappa_i$,
\begin{equation*}
u_i (a_{i}^{*}, a_{-i}^{*}) = \max_{a_{i}\in
A_{i}}\max_{a_{-i}\in \kappa_{i}} u_i (a_{i},a_{-i}) \geq
\max_{a_{i}\in A_{i}} u_{i} ( a_{i},a_{-i}^{*}).
\end{equation*}
In other words, if some $a^{*}_i$ is an optimistic best-reply to a belief $\kappa_i$, then it is a best reply to the maximizer in $\kappa_i$ given $a^{*}_i$. By the induction hypothesis, this maximizer in $\kappa_i$ is in $\PR^{m-1}_{-i}$. Finally, $\mbf{OR}^{\infty}_i = \PR^{\infty}_i$ follows from the finiteness of the game, because both procedures stop at a finite $m$. With this equivalence in mind, a version of \Mref{th:tropical_p} with $\mbf{OR}^\infty$ as the solution concept in place of Point Rationalizability can be obtained even more naturally from our `single proof for two results'.\footnote{We are extremely grateful to the anonymous associate editor for having raised our attention to this solution concept along with all the aforementioned points.}

As pointed out above, whereas it is important to recognize that Optimistic Rationalizability is actually the natural `twin' solution concept of Wald Rationalizability with the property of being equivalent to Point Rationalizability, in this paper we focus explicitly on the latter to show that this well known solution concept can be captured naturally and directly in our language, which also helps to relate it to the points made in the sections following \Sref{sec:forms_of_rationalizability}.

\subsection{Rationality in Ordinal Games}
\label{subsec:B_Rationality}

As we mentioned at the end of \Sref{subsec:motivation_results}, there is no agreement on what is the `right' notion of rationality for players in ordinal games that do not hold beliefs in the form of probability measures (and the like). This can be seen from the fact that two different notions have been proposed in the literature, each leading to different behavioral predictions when we impose them along with common belief in them. 

On one side, there is the notion of rationality as in \Mref{eq:rationality}, that we call Admissibility. This notion goes back to \citet[Definition 5, p.8]{Hillas_Samet_2014} and it is called ``Weak Dominance Rationality'' in \citet[Definition 2, p.4]{Bonanno_Tsakas_2018}. As we show in \Mref{th:tropical_a}, Admissibility and Common Belief in Admissibility epistemically characterizes the iterative elimination of actions that are B{\"o}rgers dominated (a result established in  \citet[Theorem 1, p.5]{Bonanno_Tsakas_2018} for a different framework, as we already mentioned in various instances).

On the other side, it is possible to provide a different notion of rationality as in \citet[Definition 9.3, p.417]{Bonanno_2015}, call it \emph{``Rationality$^*$''}, according to which an action $a^{*}_i$ of a player $i$ is rational$^*$ at a state if it is not the case that there exists another action $a_i$  that yields a strictly higher payoff than $a^{*}_i$ against all the action profiles of the other players that player $i$ considers possible at that state. If we focus on this notion of rationality, then \citet[Proposition 9.1, p.418]{Bonanno_2015} establishes that Rationality$^*$ and Common Belief in Rationality$^*$ is algorithmically characterized by the iterative elimination of actions that are \emph{strictly} dominated by pure actions.\footnote{See also \cite{Bonanno_2008} for an earlier result along the same lines.} \citet[Theorem 1, p.1629]{Chen_et_al_2015a}\footnote{Whose proof can be found in \citet[Section S1]{Chen_et_al_2015b}.} extend the characterization in \cite{Bonanno_2015}  to incomplete information games using possibility structures similar to our approach. Extending our characterization along the same dimension is straightforward.

Given that these notions are all based on a dominance criterion, optimism and pessimism can be seen as alternatives to the aforementioned notions of rationality, in particular in light of their solid decision-theoretic foundation.

\subsection{\texorpdfstring{Relation to \cite{Mariotti_2003}}
{Relation to Mariotti (2003)}}
\label{subsec:Mariotti_2003}

\cite{Mariotti_2003} epistemically characterizes Point Rationalizability using possibility structures, like in this contribution. However, in contrast to our approach, he focuses on 
players that choose best-replies to pure actions of the opponents without explicitly modeling---as we do---how a player chooses an action when her type considers possible multiple  actions of the opponents. 

To see the difference, consider a game $\Gamma$ with an appended possibility structure $\mfrak{P}$ and a player $i \in I$. Now, define an action $a^{*}_i \in A_i$ to be \emph{point-justifiable given}  $\widetilde{a}_{-i} \in A_{-i}$ if $a^{*}_i \in \arg \max_{a_i \in A_i}  u_i (a_i , \widetilde{a}_{-i})$.\footnote{\cite{Mariotti_2003}  uses the word ``justifiable'' instead. We use the expression ``point-justifiable'', since we employ the word ``justifiable'' in a---slightly---different way (see \Sref{sec:forms_of_rationalizability}).} Thus, action $a^{*}_i \in A_i$ is point-justifiable if the set
\begin{equation*}
	M_i (a^{*}_i) := \Set { \widetilde{a}_{-i} \in A_{-i} | %
		a^{*}_i \in \arg \max_{a_i \in A_i}  %
		u_i (a_i , \widetilde{a}_{-i})  }
\end{equation*}
is nonempty. With this definition about behavior, he proceeds by defining an epistemic event that relates the choice of player $i$'s point-justifiable actions to player $i$' types (and related possibility functions) as
\begin{equation*}
	\Mar_i := \Set { (a^{*}_i, t_i) \in A_i \times T_i | %
		M_i (a^{*}_i) \neq \emptyset , \ %
		\fob_i (t_i) \subseteq  %
		M_i (a^{*}_i) },
\end{equation*}
where it should be recalled from \Sref{sec:epistemic_static} that $\fob_i (t_i) := \proj_{A_{-i}} \pi_i (t_i)$ denotes player $i$'s first-order belief for every $t_i \in T_i$. Contrary to our approach based on the notion of optimism, $\Mar_i$ not only restricts player $i$'s behavior, but also her epistemic state. Intuitively, $\Mar_i$ can be interpreted as capturing two assumptions at once: 
\begin{itemize}
\item[i)] player $i$ chooses an action which is a best-reply to \emph{all} opponents' actions she deems possible;
\item[ii)] player $i$'s possibilities are restricted in such a way that an optimal-for-all action exists.\footnote{Formally, this would correspond to a model of decision making with incomplete preferences due to multiple point beliefs. \cite{Ziegler_Zuazo-Garin_2020} use a similar model in the realm of multiple beliefs to provide a foundation for iterated admissibility.}
\end{itemize}
Our approach, on the contrary, distinguishes assumptions about behavior and epistemic attitudes. Indeed, $\Opt_i$ is only a restriction on how player $i$ chooses an action, since in our model every type has an `optimistic' action available and no types need to be ruled out to ensure existence. 

Taking into account the discussion above, it has to be observed that the behavioral implications of both events $\Mar_i$ and $\Opt_i$ are---of course---the same: considering only types with singleton $\fob_i (t_i)$ does not change the behavioral implications of either event, but under this restriction optimistic choices are clearly point-justifiable and vice versa. However, it has to be pointed out that the goals of the two papers are different: the explicit goal of \cite{Mariotti_2003} is to epistemically characterize Point Rationalizability via possibility structures, while  our aim, rather than to provide a foundation for Point Rationalizability \emph{per se}, is to study the behavioral implications of---optimism and common belief in---optimism (and the same for pessimism) starting with an explicit formalization of these notions. 

Nevertheless, we can provide a more direct  epistemic foundation for Point Rationalizability as follows. First of all, we define the event in a possibility structure that an arbitrary player $i \in I$ has point beliefs:\footnote{That is, $\pi_i(t_i)$ being a singleton set. Within a Bayesian framework the same can be accomplished by imposing degenerate distributions as allowable beliefs.} 
\begin{equation*}
	\Deg_i := \Set { (a_i, t_i) \in A_i \times T_i | 
		\exists (a_{-i}^*, t_{-i}^*) \in A_{-i} \times T_{-i}:%
		\pi_i (t_i) = \{ (a_{-i}^*, t_{-i}^*)\} %
	}.
\end{equation*}
With this definition, the promised foundation---stated next---obtains as a corollary of \Mref{th:tropical_p}.\footnote{\Mref{cor:tropical_pp}(ii) can be established under the weaker condition of an appropriately defined \emph{degenerately belief-complete possibility structure} similar to \citet[Section 8]{Friedenberg_2019}.}
\begin{corollary}[Direct Foundation of Point Rationalizability]
	\label{cor:tropical_pp}
	Fix a game $\Gamma$.
	\begin{itemize}
		\item[i)] If $\mfrak{P}$ is an arbitrary possibility structure appended to it, then 
		\begin{equation*}
			\proj_{A} \CB^n (\Opt \cap \Deg) \subseteq \PR^{n+1},
		\end{equation*}
		for every $n \in \bN$, and 
		\begin{equation*}
			\proj_{A} \CB^\infty (\Opt \cap \Deg) \subseteq \PR^{\infty}.
		\end{equation*}
		\item[ii)] Given the universal possibility structure $\mfrak{P}^*$, 
		\begin{equation*}
			\proj_{A} \CB^n (\Opt \cap \Deg) = \PR^{n+1},
		\end{equation*}
		for every $n \in \bN$, and 
		\begin{equation*}
			\proj_{A} \CB^\infty (\Opt \cap \Deg) = \PR^{\infty}.
		\end{equation*}
	\end{itemize}
\end{corollary}

\subsection{Common Correct Belief in Optimism \emph{or} Pessimism}
\label{subsec:optimism_pessimism}

Having introduced optimism and pessimism in \Sref{sec:epistemic_static}, in  \Sref{sec:forms_of_rationalizability}, we introduced Point Rationalizability and Wald Rationalizability as the solution concepts capturing common correct belief in optimism and common correct belief in pessimism, respectively. Given this relation, it is rather natural to ask ourselves what is the solution concept related to the idea of having common correct belief of optimism \emph{or} pessimism. As a matter of fact,  it is again Wald Rationalizability that plays a crucial role, as established in the following corollary, whose proof can be found in the appendix.

\begin{corollary}
\label{cor:tropical_vn}
Fix a game $\Gamma$.
\begin{itemize}
\item[i)] If $\mfrak{P}$ is an arbitrary possibility structure appended to it, then 
\begin{equation}
\label{eq:op_m}
\proj_{A} \CB^n (\Opt \cup \Pes) \subseteq \WR^{n+1},
\end{equation}
for every $n \in \bN$, and 
\begin{equation}
\label{eq:op_infty}
\proj_{A} \CB^\infty (\Opt \cup \Pes) \subseteq \WR^{\infty}.
\end{equation}
\item[ii)] Given the universal possibility structure $\mfrak{P}^*$,
\begin{equation}
\label{eq:op_m_u}
\proj_{A} \CB^n (\Opt \cup \Pes) = \WR^{n+1},
\end{equation}
for every $n \in \bN$, and 
\begin{equation}
\label{eq:op_infty_u}
\proj_{A} \CB^{\infty} (\Opt \cup \Pes) = \WR^{\infty} .
\end{equation}
\end{itemize}
\end{corollary}

Another interpretation of \Mref{cor:tropical_vn} is one of robustness: Our characterizations in \Mref{th:tropical_wt} and \Mref{th:tropical_p} somewhat implicitly rely on the decision criterion (either $\max\max$ or $\max\min$, respectively) being transparent between the players. If players have uncertainty (and face this uncertainty under the veil of ignorance) about which of the two decision criteria are used by their opponents, then \Mref{cor:tropical_vn} shows that $\WR^{\infty}$ produces predictions that are robust to this additional uncertainty (and it does not produce superfluous predictions either).

\subsection{Common Belief vs. Common Knowledge \& Algorithmic Procedures}
\label{subsec:CB_CK}

\cite{Bonanno_Tsakas_2018} show how Admissibility  (as in \Mref{eq:rationality}---of course, in their language and terminology, where it is called ``Weak Dominance Rationality'') and Common Belief vs. Common Knowledge in Admissibility are algorithmically characterized by two different procedures. \citet[Theorem 1, p.5]{Bonanno_Tsakas_2018} (similar to our \Mref{th:tropical_a})  proves that Admissibility and Common \emph{Belief} in Admissibility is algorithmically characterized by the iterative elimination of actions that are B{\"o}rgers dominated, indeed a procedure based on the elimination of \emph{actions}. Interestingly, and clearly related to the differences between Point Rationalizability and the Wishful Thinking procedure in \cite{Yildiz_2007},  Admissibility and Common \emph{Knowledge} in Admissibility is algorithmically characterized  by an elimination of \emph{action profiles} known as Iterated Deletion of Inferior Profiles, introduced in \citet[Section 3, p.62]{Stalnaker_1994}.\footnote{\cite{Bonanno_Nehring_1998} provide a corrected proof of \citet[Theorem 3, p.63]{Stalnaker_1994}.}

\subsection{(Seemingly) Technical Assumptions}
\label{subsec:technical_assumptions}

Given that we focus on finite games, one might wonder if we need the generality provided by our topological assumption of type sets being compact Hausdorff. Indeed, this assumption might be overly general and we could work with type sets that are, for example, compact metrizable. Our characterization results rely on the existence of the universal possibility structure and \cite{Mariotti_et_al_2005} provide a canonical construction of such an object  based on the topological assumption of compact Hausdorffness.  Since our goal in this paper is to study the behavioral implications of optimism/pessimism and common belief therein and to provide a deeper understanding of ordinal games more generally, we opted to use their ready-made construction instead of providing yet another canonical construction that exactly fits our framework where the underlying space of uncertainty is finite. However, we want to stress that, even in our case, topological assumptions need to be imposed. If not, \citet[Proposition 1, p.32]{Brandenburger_2003} and \citet[Lemma 1, p.306]{Mariotti_et_al_2005} illustrate that any such construction needs to fail because it would contradict Cantor's Theorem. As pointed out in \citet[Section 11]{Brandenburger_Keisler_2006}, there is a sense in which all constructions of large structures have in common some sort of topological  assumptions.

The second parts of our characterizations results (i.e., \Mref{th:tropical_p}, \Mref{th:tropical_vn}, and \Mref{th:tropical_a}) do rely on the existence of a rich possibility structure as constructed by \cite{Mariotti_et_al_2005}. As a matter of fact, these parts of the theorems can be made stronger by only requiring a belief-complete possibility structure as defined in \citet{Brandenburger_2003} and noted in \Mref{remark:uni}. However, such a characterization would raise the question on whether such a type structure does indeed contain all hierarchies of beliefs and how the answer would depend on topological assumptions. \cite{Friedenberg_2010} addresses this question for standard type structures based on probabilistic beliefs, but it remains an open question for possibility structures. Since our formal results use the canonical construction of \cite{Mariotti_et_al_2005}, we bypass the issue by establishing the results relying on an object that contains all hierarchies of beliefs by construction.\footnote{Furthermore, because we use their construction, our results cannot just be reinterpreted as situations where players do have well-formed probabilistic beliefs, but have preferences where only the support of these beliefs matter. \citet[Section 4.2]{Mariotti_et_al_2005} discuss this point in more detail.}

\subsection{Introspection, Independence, and Knowledge in Product Structures}
\label{subsec:knowledge}

Concerning the knowledge structures in \Sref{sec:Yildiz_2007}, we can restrict our analysis to \emph{finite} structures, since the result of \cite{Yildiz_2007} (and its translation to our framework as stated in \Mref{th:tropical_wt}) does not rely on the existence of a sufficiently large (e.g., belief-complete) structure. However, we add some extra generality by allowing for knowledge structures with state spaces that do not have a product structure. Whereas in models without knowledge (and without introspection) the product structure seems---at least to us---quite natural, the product structure would impose severe restrictions on the knowledge structure. This is an implication of the Truth Axiom, which imposes cross-player restrictions as illustrated by \Mref{ex:leading_example_yildiz}. Indeed, imposing a product structure would render the players' knowledge trivial. Thus, before showing the nature of the severe restrictions mentioned above, two points are in order regarding Introspection and Independence, since they play a role in what comes next.

Regarding Introspection, it is important to observe that  is equivalent to requiring $\K_{i}(\evaluation{(a_i,t_i)})=\evaluation{(a_i,t_i)}$ for every $(s_i, t_i) \in \Psi_i \subseteq A_i \times T_i$, where $\evaluation{(a_i,t_i)}:=\Set {\omega \in \Psi | \proj_{A_i \times T_i} \omega=(a_i,t_i)}$.  

Concerning Independence, we want to highlight that, without such an assumption, conceptual problems arise concerning the interpretation of types. Indeed, in presence of Independence, a type of a player captures exactly the---interactive---knowledge that player does have, whereas this is not the case when Independence is lacking.  In particular, consider the game in \Mref{ex:leading_example_yildiz} where we already argued that $(D,L)$ cannot be played under Wishful Thinking. Now, we consider the following structure: for both players let $T_i := \{t_i\}$ with state space $\Psi := \Set {(D,R), (D,L), (U,L)}$ (type labels are omitted since they do not play an important role) and set
\begin{itemize}
	\item $\Pi_a (D, L) := \{(D, R)\}$ and $\Pi_b (D, L) := \{(U, L)\}$,
	\item $\Pi_a (D, R) := \{(D, L)\}$ and $\Pi_b (D, R) :=\{(D, R)\}$, 
	\item $\Pi_a (U, L) := \{(U, L)\}$ and $\Pi_b (U, L) := \{(D, L)\}$.
\end{itemize}
Note that, as just defined, $\Pi_i$ does not satisfy $\omega \in \Pi_i (\omega)$ or Independence, for every $i \in I$. As a result, this is not quite a knowledge structure. However, when considering the knowledge structure generated from the types only, we get a knowledge structure that satisfies introspection (for the type only) and forms a---trivial---partition. However, here we would have the state $(D,L)$ being consistent with Wishful Thinking, because, once players get informed of their \emph{own} action, they are delusional. In any case, these sort of structures are ruled out by requiring $\omega \in \Pi_i(\omega)$ and, as such, the independence assumption does not play a crucial role in \Mref{th:tropical_wt}. Indeed, \Mref{th:tropical_wt} can be strengthened to allow for knowledge structures (potentially) not satisfying Independence in the first part of the theorem and to specify that there exists a knowledge structure satisfying Independence in the second part of the theorem. Nevertheless, we opt to impose Independence on any knowledge structure considered here for the conceptual points raised above.

We can now go back to address more formally the statement previously made concerning the trivial nature of the knowledge that players would hold in a product state space. Thus,  consider a knowledge structure $\mfrak{K}$ with a product state space $\widetilde{\Psi} = \prod_{i \in I} A_i \times T_i$ and, for every player $i \in I$, let $\widetilde{\Psi}_i := \proj_{A_i \times T_i} \widetilde{\Psi}$. Additionally, fix a player $i \in I$ and, for every $E_{-i} \in \mscr{K} (\widetilde{\Psi}_{-i})$, define $\evaluation{E_{-i}} := \widetilde{\Psi}_i \times E_{-i}$ as the corresponding \emph{interactive event (for player $i$)}. Interestingly, given our assumption that the knowledge structure satisfies Introspection and Independence, the following remark states that the only interactive event a player knows is the full state space.

\begin{remark}
Given a game $\Gamma$ and an appended knowledge structure $\mfrak{K}$ such that $\widetilde{\Psi} = \prod_{i \in I} A_i \times T_i$. For every player $i \in i$,
	\begin{equation*}
\mbb{K}_i ( \evaluation{E_{-i}}) \neq \emptyset \Llra E_{-i} = \widetilde{\Psi}_{-i},
	\end{equation*}
for every event $E_{-i} \in \mscr{K} \big( \widetilde{\Psi}_{-i} \big)$. 
\end{remark}

For one, just note that $E_{-i} = \widetilde{\Psi}_{-i}$ says that $\evaluation{E_{-i}} = \widetilde{\Psi}$ and then $\mbb{K}_i ( \evaluation{E_{-i}}) = \mbb{K}_i \big( \widetilde{\Psi} \big) = \widetilde{\Psi}$.  For the converse, assume that $E_{-i} \neq \widetilde{\Psi}_{-i}$, i.e., $E_{-i} \subsetneq \widetilde{\Psi}_{-i}$. By Introspection and Independence, every partition cell for a given $\omega=(\omega_i, \omega_{-i}) \in \widetilde{\Psi}$ is of the form $\Pi_i (\omega) = \{\omega_i\} \times \widehat{E}^{t_{i}^{\omega}}_{-i}$, for a $t_{i}^{\omega} := \proj_{ T_i} \omega$ and a $\widehat{E}^{t_{i}^{\omega} }_{-i} \in \mscr{K} \big( \widetilde{\Psi}_{-i} \big)$. Thus, we can index the partition cells by $\omega_i$ only, which implies, given the definition of a partition, that we also need $\bigcup_{\omega_i \in \widetilde{\Psi}_i} \Pi_i (\omega_i) = \widetilde{\Psi}$. This, in turn, implies that every cell needs to be of the form $\Pi_i (\omega) = \{ \omega_i \} \times \widetilde{\Psi}_{-i}$, i.e. they are cylinder sets. Thus, for every $\omega \in \widetilde{\Psi}$, $\Pi_i (\omega) = \{ \omega_i \} \times \widetilde{\Psi}_{-i} \nsubseteq \evaluation{E_{-i}}$, i.e., $\mbb{K}_i ( \evaluation{E_{-i}}) = \emptyset$.

Although the construction of a product structure mentioned above is somewhat standard (see---for example--- \citet[p. 429]{Zamir_2009}), the  remark just stated about product structures is---even if simple---new to the best of our knowledge, where---of course---our insistence on requiring Introspection and Independence comes from the very fact that we want to relate knowledge structures to our possibility structures.

\appendix

\section*{Appendix}

\section{Proofs}
\label{app:proofs}

\subsection{\texorpdfstring{Proofs of \Sref{sec:epistemic_static}}{Proofs of Section 2}}
\label{subsec:proofs_static}

Given an arbitrary topological space $X$, $\mscr{B} (X)$ denotes its Borel $\sigma$-algebra and $\abs{X}$ its cardinality. Given our topological assumptions spelled out in \Sref{sec:epistemic_static}, we can state the following remark.

\begin{remark}[Measurability]
\label{rem:events}
If $X$ is compact Hausdorff, then $\mscr{K} (X) \subseteq \mscr{B} (X)$. 
\end{remark}

We provide a unique proof for all the results in \Sref{sec:epistemic_static}. For this and to ease notation, henceforth, we let $(\rho, \Att) \in \{ \Rounds{ \obr, \Opt}, \Rounds{ \pbr, \Pes} \}$.

\begin{remark}
\label{rem:CB_reformulation}
For every $(\rho, \Att) \in \{ \Rounds{ \obr, \Opt}, \Rounds{ \pbr, \Pes} \}$ and $i \in I$,
\begin{equation}
\label{eq:CB_reformulation}
\proj_{\Omega_i} \CB^n (\Att) = %
\proj_{\Omega_i} \CB^{n-1} (\Att) \cap \Bel_i %
\Rounds{ \proj_{\Omega_{-i} }\CB^{n-1} (\Att) },
\end{equation}
for every $n \geq 1$.
\end{remark}

\begin{proof}[Proof of \Mref{prop:measurability_static}]
Fix a possibility structure $\mfrak{P}$ with state space $\Omega$ appended to a game $\Gamma$, a  $(\rho, \Att) \in \{ \Rounds{ \obr, \Opt}, \Rounds{ \pbr, \Pes} \}$, and a player $i \in I$. 

\begin{itemize}[leftmargin=*]
\item[$i)$] We now proceed with the proof of part (i). In light of \Mref{rem:CB_reformulation}, we are going to establish the truth of \Mref{eq:CB_reformulation}. To do so, we proceed by induction on $n \in \bN$.
\begin{itemize}[leftmargin=*]
\item[$\bullet$] ($n = 0$) First of all, notice that, since
\begin{equation*}
\proj_{\Omega_i} \CB^0 (\Att) = \Att_i = %
\bigcup_{a_i \in A_i} \Squares{ \{ a_i \} \times \proj_{T_i} %
\Rounds{\Att_i \cap (\{a_i\} \times T_i)} } ,
\end{equation*}
we have to prove that $\proj_{T_i} \Rounds{\Att_i \cap (\{a^{*}_i\} \times T_i)}$ is closed for an arbitrary $a^{*}_i \in A_i$. Now, we have that 
\begin{equation*}
\proj_{T_i} \Rounds{\Att_i \cap (\{a^{*}_i\} \times T_i)} = %
\pi^{-1}_i \Rounds{ \Set{ \xi_i \in \mscr{K} (A_{-i} \times T_{-i}) | %
a^{*}_i \in \rho_i \Rounds {\proj_{A_{-i}} \xi_i} }}.
\end{equation*}
Thus, since $\pi_i$ is continuous by assumption, we simply have to show that the set 
\begin{equation*}
\Set{ \xi_i \in \mscr{K} (A_{-i} \times T_{-i}) | %
a^{*}_i \in \rho_i \Rounds {\proj_{A_{-i}} \xi_i} }
\end{equation*}
is closed. Let $(\widetilde{\xi}^{\ell}_i)^{\ell \in \bN} \subseteq \Omega_{-i}$ be a sequence such that $a^{*}_i \in \rho_i \Rounds {\proj_{A_{-i}} \widetilde{\xi}^{\ell}_i}$ for every $\ell \in \bN$ and assume that $\widetilde{\xi}^{\ell}_i \to \widetilde{\xi}_i$. Thus, we need to prove that $a^{*}_i \in \rho_i \Rounds {\proj_{A_{-i}} \widetilde{\xi}_i}$. Now, for every $\ell \in \bN$, $\proj_{A_{-i}} \widetilde{\xi}^{\ell}_i \subseteq A_{-i}$ with $A_{-i}$ finite and---by assumption---endowed with the discrete topology. Also, recall that convergence of a sequence in the discrete topology means that  there exists a $\widehat{k} \in \bN$ such that, for every $m > \widehat{k}$, $\proj_{A_{-i}} \widetilde{\xi}^{\widehat{k}}_i = \proj_{A_{-i}} \widetilde{\xi}^{m}_i$. Thus, we have---\emph{a fortiori}---also that $\proj_{A_{-i}} \widetilde{\xi}_i = \proj_{A_{-i}} \widetilde{\xi}^{\widehat{k}}_i$. Hence, it follows that $a^{*}_i \in \rho_i \Rounds {\proj_{A_{-i}} \widetilde{\xi}_i}$.
\item[$\bullet$] ($n \geq 1$) Assume the result holds for $n \in \bN$. Thus, we have to prove that $\CB^{n+1} (\Att) \in \mscr{K} (\Omega)$. Let $i \in I$ be arbitrary and, focusing on \Mref{eq:CB_reformulation}, observe that we have  $\proj_{\Omega_i} \CB^n (\Att) \in \mscr{K} (\Omega)$, from the induction hypothesis. Thus, it remains to prove that $\Bel_i \Rounds{\proj_{\Omega_{-i}} \CB^n (\Att)} \in \mscr{K} (\Omega_i)$. Now, notice that 
\begin{equation*}
\Bel_i \Rounds{\proj_{\Omega_{-i}} \CB^n (\Att)} = %
\pi^{-1}_i \Rounds { \Set {\xi_i \in \mscr{K} (A_{-i} \times T_{-i}) | %
\xi_i \subseteq \proj_{\Omega_{-i}} \CB^n (\Att)} }.
\end{equation*}
Thus, since $\pi_i$ is continuous by assumption, we simply have to show that the set 
\begin{equation*}
\Set {\xi_i \in \mscr{K} (A_{-i} \times T_{-i}) | %
\xi_i \subseteq \proj_{\Omega_{-i}} \CB^n (\Att)}
\end{equation*}
is closed, which is immediately established by noticing that  $\proj_{\Omega_{-i}} \CB^n (\Att)$ is closed from the induction hypothesis.\footnote{This step can be alternatively proven in a more explicit fashion by showing that $\Bel_i (E_{-i})$ is closed whenever $E_{-i}$ is closed by employing a convergence argument in the Hausdorff metric. We are grateful to an anonymous referee for having pointed out this alternative path.} 
\end{itemize}

\item[$ii)$] Regarding part (ii), the result follows immediately from part (i), since an intersection of closed sets is a closed set.

\end{itemize}

\noindent This establishes the result.
\end{proof}

\subsection{\texorpdfstring{Proofs of \Sref{sec:forms_of_rationalizability}}{Proofs of Section 3}}
\label{subsec:proofs_algo}

For the purpose of the proofs contained in this section, we rewrite \Mref{eq:p_def} and \Mref{eq:v_def} as follows
\begin{equation*}
\PR^{m}_{i}  := %
\Set { a^{*}_i \in \PR^{m-1}_{i} | %
\begin{array}{l}
\exists \kappa_i \in \mscr{K} (A_{-i}) \ 
\exists a^{*}_{-i} \in \PR^{m-1}_{-i} : \\%
1.\  \kappa_i = \{ a^{*}_{-i} \} , \\
2.\  a^{*}_i \in \obr_i (\kappa_i)
\end{array}
},\\
\end{equation*}
and
\begin{equation*}
\WR^{m}_{i}  := %
\Set { a^{*}_i \in \WR^{m-1}_{i} | %
\begin{array}{l}
\exists \kappa_i \in \mscr{K} (A_{-i}) \ 
\exists \widetilde{A}_{-i} \subseteq \WR^{m-1}_{-i} : \\%
1.\  \kappa_i = \widetilde{A}_{-i}  , \\
2.\  a^{*}_i \in \pbr_i (\kappa_i)
\end{array}
}.%
\end{equation*}
These formulations---clearly equivalent to \Mref{eq:p_def} and \Mref{eq:v_def}---make more perspicuous the nature of the proof that follows that, as for the results in the previous section, leads to a \emph{unique} proof for all the results in \Sref{sec:forms_of_rationalizability}. Thus, in the following---joint---proof, we  let 
\begin{equation*}
(\AR, \rho, \Att) \in %
\{ \Rounds{ \PR, \obr, \Opt}, %
	\Rounds{ \WR, \pbr, \Pes} \}. %
\end{equation*}

We divide the proof of \Mref{th:tropical_p}/\Mref{th:tropical_vn} in two parts for clarity of exposition. Of course we start from part (i) and then move to part (ii).
Concerning part (i), we need additional notation. That is, given an action-type pair $(a^{*}_i , \widetilde{t}_i) \in A_i \times T_i$, we let $\kappa^{\widetilde{t}_i}_i \in \mscr{K} (A_{-i})$ be defined as
\begin{equation}
\label{eq:kappa}
\kappa^{\widetilde{t}_i}_i :=
\begin{cases}
\displaystyle \{a^{*}_{-i}\} : a^{*}_{-i} \in %
\arg \max_{a_{-i} \in \fob_i (\widetilde{t}_i)} %
u_i (a^{*}_i, a_{-i}), & %
\text{ if } (\AR, \rho, \Att) = %
(\PR, \obr, \Opt), \\
\displaystyle  \fob_i (\widetilde{t}_i)  , & %
\text{ otherwise, } 
\end{cases}
\end{equation}
where in both cases we have by construction that $\kappa^{\widetilde{t}_i}_i \subseteq \fob_i (\widetilde{t}_i)$.\footnote{This definition directly takes care of the difference of Point and Optimistic Rationalizability as discussed in \Sref{subsec:OR}.}  

\begin{proof}[Proof of \Mref{th:tropical_p}/\Mref{th:tropical_vn}(i)]
We divide the proof in two parts. We proceed by proving \Mref{eq:p_m}/\Mref{eq:v_m} first and then move to prove \Mref{eq:p_infty}/\Mref{eq:v_infty}. Fix a tuple $(\AR, \rho, \Att)$.
\begin{itemize}[leftmargin=*]
\item Regarding the proof of \Mref{eq:p_m}/\Mref{eq:v_m}, we proceed by  induction on $n \in \bN$.
\begin{itemize}[leftmargin=*]
\item ($n = 0$) Let $(a^{*}, \widetilde{t}) \in \Att$ and $i \in I$ be arbitrary. Let $\kappa^{\widetilde{t}_i}_i \in \mscr{K} (A_{-i})$ be defined as in \Mref{eq:kappa}. From our assumption, $a^{*}_i \in \rho_i \big( \kappa^{\widetilde{t}_i}_i \big)$. Hence, $a^{*}_i \in \AR^{1}_i$. 
\item ($n \geq 1$) Fix an $n \geq 1$, assume the result holds for $n -1$, and let $(a^{*}, \widetilde{t}) \in \CB^n (\Att)$ and $i \in I$ be arbitrary. Hence, $\pi_i (\widetilde{t}_i) \subseteq \proj_{\Omega_{-i}} \CB^{n-1} (\Att)$. Let $\kappa^{\widetilde{t}_i}_i \in \mscr{K} (A_{-i})$ be defined as in \Mref{eq:kappa}. From the induction hypothesis, $\kappa^{\widetilde{t}_i}_i \subseteq \AR^{n}_{-i}$. Thus,  since---\emph{a fortiori}---we have that $(a^{*}_i, \widetilde{t}_i)  \in \Att_i$, it is the case that $a^{*}_i \in \rho_i \big( \kappa^{\widetilde{t}_i}_i \big)$. Hence, it follows that $a^{*}_i \in \AR^{n+1}_i$, because $\kappa^{\widetilde{t}_i}_i \subseteq \AR^{n}_{-i}$.
\end{itemize}
\item \Mref{eq:p_infty}/\Mref{eq:v_infty} immediately follow from  \Mref{eq:p_m}/\Mref{eq:v_m}, the finiteness assumption, and the nonemptiness of the solution concepts. \qedhere
\end{itemize}
\end{proof}

\begin{proof}[Proof of \Mref{th:tropical_p}/\Mref{th:tropical_vn}(ii)]
Let $\mfrak{P}^{*}$ be the universal possibility structure. Fix a tuple $(\AR, \rho, \Att)$.
\begin{itemize}[leftmargin=*]
\item We now prove \Mref{eq:p_bc_m}/\Mref{eq:v_bc_m}. Clearly, one side of the result has already been established in the proof of part (i). Thus, we establish the other side of the result by proceeding again by induction on $n \in \bN$.
\begin{itemize}[leftmargin=*]
\item ($n = 0$) Fix a profile of actions $a^{*} \in \AR^{1}$ and let $i \in I$ be arbitrary. Then there exists a  $\kappa_{i} \in  \mscr{K} (A_{-i})$ such that $\displaystyle a^{*}_i \in \rho_i (\kappa_i)$. From the belief-completeness of $\mfrak{P}^*$, there exists a type $\widetilde{t}_i \in T^{*}_i$ such that $\pi_i (\widetilde{t}_i) = \kappa_i \times T^{*}_{-i}$. Thus, it follows that $(a^{*}_i , \widetilde{t}_i) \in \Att_i$ by construction. Since the player $i$ was chosen arbitrarily, the result follows.
\item ($n \geq 1$) Fix an $n \geq 1$, assume the result holds for $n-1$, and fix a profile of actions $a^{*} \in \AR^{n+1}$. Let $i \in I$ be arbitrary. Then there exists a $\kappa_i \in \mscr{K} (A_{-i})$ with $\kappa_i \subseteq \AR^{n}_{-i}$ such that $a^{*}_i \in \rho_i (\kappa_i)$. From the induction hypothesis, for every $a_{-i} \in \kappa_i$ there exists a type $t^{a_{-i}}_{-i} \in T_{-i}$ such that $(a_{-i}, t^{a_{-i}}_{-i}) \in \proj_{\Omega_{-i}} \CB^{n-1}(\Att)$. Hence, from the belief-completeness of $\mfrak{P}^*$, there exists a type $\widetilde{t}_i \in T^{*}_i$ such that 
\begin{equation*}
\pi_i (\widetilde{t}_i) := %
\Set { \Rounds{a_{-i}, t^{a_{-i}}_{-i}} \in A_{-i} \times T^{*}_{-i} | a_{-i} \in \kappa_i }
\end{equation*}
and---by construction---we have that $(a^{*}_i, \widetilde{t}_i) \in \proj_i \CB^n (\Att)$. Since player $i$ was chosen arbitrarily, the result follows. 
\end{itemize}
\item We now prove \Mref{eq:p_bc_infty}/\Mref{eq:v_bc_infty}, where---again---we already established one side in the proof above. Thus, first of all, observe that $\CB^{\infty} (\Att) \neq \emptyset$. This is a consequence of the fact that $\AR^n \neq \emptyset$ for every $n \in \bN$ and that $T$ is compact Hausdorff by assumption. Hence, $(\CB^{m} (\Att))_{m \geq 0}$ is a nested family of nonempty closed sets having the finite intersection property.  Let $\underline{n} := \min \Set { n \in \bN | \AR^n = \AR^{n+1} = \AR^\infty }$. Let $a^* \in \AR^{\underline{n}} = \AR^\infty$ be arbitrary. Let
\begin{equation*}
M^\ell (\underline{n}, a^* ) := 
\begin{cases}
\{a^* \} \times T^{*} , & \text{ if } \underline{n} = 0, \\
\CB^{{\underline{n} - 1} + \ell} (\Att) \cap (\{a^* \} \times T^{*} ), & \text{ otherwise},
\end{cases}
\end{equation*}
for every $\ell \geq 0$. Notice that this definition induces a sequence of sets. Since every $M^\ell (\underline{n} , a^*)$ is nonempty and closed and the sequence of sets is decreasing, it has the finite intersection property. Hence, there exists a $t^{*} \in T^{*}$ such that $(a^{*},  t^*) \in \bigcap_{\ell \geq 0} M^\ell (\underline{n} , a^*) \subseteq \CB^\infty (\Att)$.
\end{itemize}

\noindent This completes the proof of part (ii). \qedhere
\end{proof}

\begin{proof}[Proof of \Mref{prop:P-V}]
We proceed by induction on $n \in \bN$.
\begin{itemize}[leftmargin=*]
\item ($n = 0$) Trivial.
\item ($n \geq 1)$ Fix an $n \geq 1$ and assume the result holds for $n-1$. Let $a^{*} \in \PR^{n}$ and $i \in I$ be arbitrary. Hence, there exists a $a_{-i} \in \PR^{n-1}_{-i}$ such that $a^{*}_i \in \obr_i (\kappa_i)$, with $\kappa_i := \{ a_{-i} \}$. Let $\widehat{A}_{-i} := \kappa_i$. Then, \emph{a fortiori} also $a^{*}_i \in \pbr_i (\kappa_i)$.
\end{itemize}
This completes the proof. \qedhere
\end{proof}

\subsection{\texorpdfstring{Proofs of \Sref{sec:Yildiz_2007}}{Proofs of Section 4}}
\label{subapp:proofs_Y}

\begin{proof}[Proof of \Mref{prop:Y-P}]
We fix a game $\Gamma$ and proceed by induction on $n \in \bN$.
\begin{itemize}[leftmargin=*]
\item ($n = 0$) Trivial.
\item ($n \geq 1)$ Fix an $n \geq 1$ and assume the result holds for $n-1$. Let $a^{*} \in  \YR^{n}$ and $i \in I$ be arbitrary. Hence, there exists an $a_{-i} \in A_{-i}$ such that  $(a_i^*, a_{-i}) \in  \YR^{n-1} \subseteq \PR^{n-1}$ from the induction hypothesis and $a^{*}_i \in \obr_i (\{a_{-i}\})$. Thus, the conclusion follows.
\end{itemize}
This completes the proof. \qedhere
\end{proof}

The proof of \Mref{th:tropical_wt} can be easily obtained by `translating' the proof of \citet[Proposition 1, p.327]{Yildiz_2007} in \citet[pp.341-342]{Yildiz_2007}  to our framework, with the understanding that the event $W_i$ in \cite{Yildiz_2007} is equivalent to our $\Opt_i$ (in particular, as defined in \Sref{foot:opt_Yildiz}). Some care has to be taken when constructing the relevant state space by adding the appropriate types that are explicit in our framework. Furthermore, the statement in \citet[Proposition 1, p.327]{Yildiz_2007}  is slightly different from ours since  a (possibly distinct) knowledge structure is constructed for every step $m\geq 0$. To obtain our statement, one can just take the union across all the knowledge structures as the relevant knowledge structure. The reason the arguments in \citet[pp.341-342]{Yildiz_2007} can be translated in the present setting is that, for most of the arguments therein, the Bayesian framework with probabilistic beliefs employed does not play a crucial role. In particular, many steps rely on the existence of point-beliefs, which we do have in our non-probabilistic framework too. Only, \citet[Lemma 2, p.341]{Yildiz_2007}  uses properties of expectations: as a result, we need to prove the corresponding result in our framework (stated below) in a slightly different fashion.

\begin{lemma}
	For every $F \subseteq \Psi$,  $i \in I$, and  $\widehat{a} :=  (\widehat{a}_i, \widehat{a}_{-i}) \in \proj_A \K_i (F)  \cap \Opt_i$, there exists a $(\widehat{a}_i, a_{-i}) \in \proj_A F$ such that:
	\begin{itemize} 
	\item[$(i)$] $\widehat{a}_i \in \obr_i (\{a_{-i}\})$,
	\item[($ii$)]   $u_i (\widehat{a}_i, a_{-i}) \geq \max_{a_i \in A_i} u_i (a_i, \widehat{a}_{-i})$.	
	\end{itemize}
\end{lemma}

\begin{proof}
Let $\widehat{a} :=  (\widehat{a}_i, \widehat{a}_{-i}) \in \proj_A \K_i(F) \cap \Opt_i$ be arbitrary, with $\omega$ be a corresponding state in $\K_i (F) \cap \Opt_i$. By Introspection, $\{ \widehat{a}_i \}= \proj_{A_i} \Pi_i (\omega)$. Therefore, for every $\omega' \in  \Pi_i (\omega)$, we need to have the same action $\widehat{a}_i$ prescribed for player $i$. Thus, let 
\begin{equation*}
a_{-i}^\omega \in \argmax_{a_{-i} \in \proj_{A_{-i}} \Pi_i (\omega)} u_i(\widehat{a}_i, a_{-i})
\end{equation*}
and note that  $(\widehat{a}_i, a_{-i}^\omega) \in \argmax_{a \in \proj_A \Pi_i (\omega)} u_i(a)$. Now, since $\omega \in \K_i (F)$, we have that $\Pi_i (\omega) \subseteq F$. Therefore, $\proj_{A} \Pi_i(\omega) \subseteq \proj_{A} F$. Hence, $(\widehat{a}_i, a_{-i}^\omega) \in \proj_{A} F$. Thus, by Wishful Thinking, $\widehat{a}_i \in \argmax_{a_i \in A_i} u_i (a_i, a_{-i}^\omega)$. Therefore, $\widehat{a}_i \in \obr_i (\{a_{-i}^\omega\})$. Finally, since $\omega \in \Pi_i (\omega)$, we know that $\widehat{a}_{-i} \in \proj_{A_{-i}} \Pi_i(\omega)$. Thus,
\begin{equation*}
u_i(\widehat{a}_i, a_{-i}^\omega) = %
\max_{a_i \in A_i} u_i(a_i, a_{-i}^\omega) = %
\max_{a_i \in A_i} \max_{a_{-i} \in \proj_{A_{-i}} \Pi_i(\omega)} u_i(a_i, a_{-i})\geq%
\max_{a_i \in A_i} u_i (a_i, \widehat{a}_{-i}). \qedhere
\end{equation*}
\end{proof}

\subsection{\texorpdfstring{Proofs of \Sref{sec:B-dominance}}{Proofs of Section 5}}
\label{subapp:proofs_B}

Regarding the measurability as in \Mref{prop:measurability_A} of $\CB^m (\Ad)$, for every $m \geq 0$, and $\ACBA$, the proofs in \Sref{subsec:proofs_static} apply \emph{verbatim} with $(\rho, \Att) = \Rounds{ \rho^B, \Ad}$, where the same applies to the proof of \Mref{th:tropical_a} as proved in \Sref{subsec:proofs_algo} with $(\AR, \rho, \Att) = \Rounds{ \BR, \rho^B, \Ad}$ and  $\kappa^{\widetilde{t}_i}_i \in \mscr{K} (A_{-i})$ be defined as $\kappa^{\widetilde{t}_i}_i := \fob_i (\widetilde{t}_i) \subseteq A_{-i}$.

\begin{proof}[Proof of \Mref{prop:P_B}]
We fix a game $\Gamma$ and proceed by induction on $n \in \bN$.
\begin{itemize}[leftmargin=*]
\item ($n = 0$) Trivial.
\item ($n \geq 1)$ Fix an $n \geq 1$ and assume the result holds for $n-1$. Let $a^{*} \in \PR^{n}$ and $i \in I$ be arbitrary. Hence, there exists a $\kappa_i \in \mscr{K} (A_{-i})$ such that $a^{*}_i \in \obr_i (\kappa_i)$, with $\kappa_{i} := \{ \widetilde{a}_{-i}\}$  for a $\widetilde{a}_{-i} \in \PR^{n-1}_{-i}$. From the induction hypothesis, $\widetilde{a}_{-i} \in \BR^{n-1}_i$. Hence, $a^{*}_i \in \bbr_i (\kappa_i)$.
\end{itemize}
This completes the proof. \qedhere
\end{proof}

\begin{proof}[Proof of \Mref{prop:generic_V_B}]
In generic games, given an arbitrary player $i \in I$, an action $a_i \in A_i$ is B-dominated if and only if it is strictly dominated by a pure action.\footnote{See \citet[Footnote 5, p.1884]{Weinstein_2016}.} Hence, this establishes the result, for every $n \in \bN$.
\end{proof}

\subsection{\texorpdfstring{Proofs of \Sref{sec:discussion}}{Proofs of Section 7}}
\label{subsec:proofs_discussion}

As usual, we divide the proof of \Mref{cor:tropical_vn}  in two parts for clarity of exposition by starting from part (i) to then moving to part (ii).

\begin{proof}[Proof of \Mref{cor:tropical_vn}(i)]
We divide the proof in two parts. We proceed by proving \Mref{eq:op_m} first and then move to prove \Mref{eq:op_infty}. 
\begin{itemize}[leftmargin=*]
\item Regarding the proof of \Mref{eq:op_m}, we proceed by  induction on $n \in \bN$.
\begin{itemize}[leftmargin=*]
\item ($n = 0$) Let $(a^{*}, \widetilde{t}) \in \Opt \cup \Pes$ and $i \in I$ be arbitrary. Let $\kappa^{\widetilde{t}_i}_i \in \mscr{K} (A_{-i})$ be defined as $\kappa^{\widetilde{t}_i}_i := \{a^{*}_{-i}\}$ such that
\begin{equation*}
a^{*}_{-i} \in \arg \max_{a_{-i} \in \fob_i (\widetilde{t}_i)} %
u_i (a^{*}_i, a_{-i}). 
\end{equation*}
From our assumption, $a^{*}_i \in \obr_i \big( \kappa^{\widetilde{t}_i}_i \big) \cup \pbr_i \big( \kappa^{\widetilde{t}_i}_i \big)$. Since $\pbr_i (\kappa_i) = \obr_i (\kappa_i)$ for $\kappa_i$ singleton, it follows that $a^{*}_i \in \pbr_i \big( \kappa^{\widetilde{t}_i}_i \big)$. Hence, $a^{*}_i \in \WR^{1}_i$. 
\item ($n \geq 1$) Fix an $n \geq 1$, assume the result holds for $n -1$, and let $(a^{*}, \widetilde{t}) \in \CB^n (\Opt \cup \Pes)$ and $i \in I$ be arbitrary. Hence, $\pi_i (\widetilde{t}_i) \subseteq \proj_{\Omega_{-i}} \CB^{n-1} (\Opt \cup \Pes)$. Let $\kappa^{\widetilde{t}_i}_i \in \mscr{K} (A_{-i})$ be defined as $\kappa^{\widetilde{t}_i}_i := \{a^{*}_{-i}\}$ such that
\begin{equation*}
a^{*}_{-i} \in \arg \max_{a_{-i} \in \fob_i (\widetilde{t}_i)} %
u_i (a^{*}_i, a_{-i}). 
\end{equation*}
From the induction hypothesis, $\kappa^{\widetilde{t}_i}_i \subseteq \WR^{n}_{-i}$. Thus,  since---\emph{a fortiori}---we have that $(a^{*}_i, \widetilde{t}_i)  \in \Opt_i \cup \Pes_i$, it is the case that $a^{*}_i \in \obr_i \big( \kappa^{\widetilde{t}_i}_i \big) \cup \pbr_i \big( \kappa^{\widetilde{t}_i}_i \big)$, which implies that $a^{*}_i \in \pbr_i \big( \kappa^{\widetilde{t}_i}_i \big)$ from the equivalence of $\pbr_i (\kappa_i)$ and $\obr_i (\kappa_i)$ for $\kappa_i$ singleton. Hence, it follows that $a^{*}_i \in \WR^{n+1}_i$, because $\kappa^{\widetilde{t}_i}_i \subseteq \WR^{n}_{-i}$.
\end{itemize}
\item \Mref{eq:op_infty} immediately follows from  \Mref{eq:op_m}, the finiteness assumption, and the nonemptiness of the solution concepts. \qedhere
\end{itemize}
\end{proof}

\begin{proof}[Proof of \Mref{cor:tropical_vn}(ii)]
Let $\mfrak{P}^{*}$ be the universal possibility structure.
\begin{itemize}[leftmargin=*]
\item We now prove \Mref{eq:op_m_u}. Clearly, one side of the result has already been established in the proof of part (i). Thus, we establish the other side of the result by proceeding again by induction on $n \in \bN$.
\begin{itemize}[leftmargin=*]
\item ($n = 0$) Fix a profile of actions $a^{*} \in \WR^{1}$ and let $i \in I$ be arbitrary. Then there exists a  $\kappa_{i} \in  \mscr{K} (A_{-i})$ such that $\displaystyle a^{*}_i \in \pbr_i (\kappa_i)$. From the belief-completeness of $\mfrak{P}^*$, there exists a type $\widetilde{t}_i \in T^{*}_i$ such that $\pi_i (\widetilde{t}_i) = \kappa_i \times T^{*}_{-i}$. Thus, it follows that $(a^{*}_i , \widetilde{t}_i) \in \Pes_i$ by construction and we have---\emph{a fortiori}---that $(a^{*}_i , \widetilde{t}_i) \in \Opt_i \cup \Pes_i$. Since the player $i$ was chosen arbitrarily, the result follows.
\item ($n \geq 1$) Fix an $n \geq 1$, assume the result holds for $n-1$, and fix a profile of actions $a^{*} \in \WR^{n+1}$. Let $i \in I$ be arbitrary. Then there exists a $\kappa_i \in \mscr{K} (A_{-i})$ with $\kappa_i \subseteq \WR^{n}_{-i}$ such that $a^{*}_i \in \pbr_i (\kappa_i)$. From the induction hypothesis, for every $a_{-i} \in \kappa_i$ there exists a type $t^{a_{-i}}_{-i} \in T_{-i}$ such that $(a_{-i}, t^{a_{-i}}_{-i}) \in \proj_{\Omega_{-i}} \CB^{n-1}(\Opt \cup \Pes)$. Hence, from the belief-completeness of $\mfrak{P}^*$, there exists a type $\widetilde{t}_i \in T^{*}_i$ such that 
\begin{equation*}
\pi_i (\widetilde{t}_i) := %
\Set { \Rounds{a_{-i}, t^{a_{-i}}_{-i}} \in A_{-i} \times T^{*}_{-i} | a_{-i} \in \kappa_i }
\end{equation*}
and---by construction---we have that $(a^{*}_i, \widetilde{t}_i) \in \proj_i \CB^n (\Opt \cup \Pes)$. Since player $i$ was chosen arbitrarily, the result follows. 
\end{itemize}
\item We now prove \Mref{eq:op_infty_u}, where we already established one side above. Thus, first of all, observe that $\CB^{\infty} (\Opt \cup \Pes) \neq \emptyset$. This is a consequence of \Mref{eq:op_m_u}, the fact that  $\WR^n \neq \emptyset$ for every $n \in \bN$, and the assumptions that the games are finite and that $T$ is compact Hausdorff. Hence, $(\CB^{m} (\Opt \cup \Pes))_{m \geq 0}$ is a nested family of nonempty closed sets with the finite intersection property.  Let $\underline{n} := \min \Set { n \in \bN | \WR^n = \WR^{n+1} = \WR^\infty }$. Let $a^* \in \WR^{\underline{n}} = \WR^\infty$ be arbitrary and let
\begin{equation*}
M^\ell (\underline{n}, a^* ) := 
\begin{cases}
\{a^* \} \times T^{*} , & \text{ if } \underline{n} = 0, \\
\CB^{{\underline{n} - 1} + \ell} (\Opt \cup \Pes) \cap (\{a^* \} \times T^{*} ), & \text{ otherwise},
\end{cases}
\end{equation*}
for every $\ell \geq 0$. Notice that this definition induces a sequence of sets. Since every $M^\ell (\underline{n} , a^*)$ is nonempty and closed and the sequence of sets is decreasing, it has the finite intersection property. Hence, there exists a $t^{*} \in T^{*}$ such that $(a^{*},  t^*) \in \bigcap_{\ell \geq 0} M^\ell (\underline{n} , a^*) \subseteq \CB^\infty (\Opt \cup \Pes)$.
\end{itemize}

\noindent This completes the proof of part (ii). \qedhere
\end{proof}

%
\hypersetup{colorlinks=true,linkcolor=green!50!black}
\phantomsection
\addcontentsline{toc}{section}{References}

\bibliography{Biblio_Tropical_Players.bib}{}

\begin{thebibliography}{68}
\providecommand{\natexlab}[1]{#1}
\expandafter\ifx\csname urlstyle\endcsname\relax
  \providecommand{\doi}[1]{doi:\discretionary{}{}{}#1}\else
  \providecommand{\doi}{doi:\discretionary{}{}{}\begingroup
  \urlstyle{rm}\Url}\fi

\bibitem[{Arrow \& Hurwicz(1977)}]{Arrow_Hurwicz_1977}
\textsc{Arrow, K.J., Hurwicz, L.} (1977).
\newblock An Optimality Criterion for Decision-Making under Ignorance.
\newblock In \emph{Studies in Resource Allocation Processes} (K.J. Arrow,
  L.~Hurwicz, editors). Cambridge University Press, Cambridge.

\bibitem[{Aumann(1976)}]{Aumann_1976}
\textsc{Aumann, R.J.} (1976).
\newblock Agreeing to Disagree.
\newblock \emph{The Annals of Statistics}, \textbf{4}, 1236--1239.

\bibitem[{Aumann(1999)}]{Aumann_1999a}
---{}---{}--- (1999).
\newblock Interactive Epistemology I: Knowledge.
\newblock \emph{International Journal of Game Theory}, \textbf{28}, 263--300.

\bibitem[{Bach \& Perea(2020)}]{Bach_Perea_2020}
\textsc{Bach, C.W., Perea, A.} (2020).
\newblock Two Definitions of Correlated Equilibrium.
\newblock \emph{Journal of Mathematical Economics}, \textbf{90}, 12--24.

\bibitem[{Battigalli \& Bonanno(1999)}]{Battigalli_Bonanno_1999}
\textsc{Battigalli, P., Bonanno, G.} (1999).
\newblock Recent Results on Belief, Knowledge and the Epistemic Foundations of
  Game Theory.
\newblock \emph{Research in Economics}, \textbf{53}, 149--225.

\bibitem[{Battigalli et~al.(2016)Battigalli, Cerreia-Vioglio, Maccheroni, \&
  Marinacci}]{Battigalli_et_al_2016}
\textsc{Battigalli, P., Cerreia-Vioglio, S., Maccheroni, F., Marinacci, M.}
  (2016).
\newblock A Note on Comparative Ambiguity Aversion and Justifiability.
\newblock \emph{Econometrica}, \textbf{84}, 1903--1916.

\bibitem[{Battigalli et~al.(Work in Progress)Battigalli, Friedenberg, \&
  Siniscalchi}]{Battigalli_et_al_forthcoming}
\textsc{Battigalli, P., Friedenberg, A., Siniscalchi, M.} (Work in Progress).
\newblock Epistemic Game Theory: Reasoning about Strategic Uncertainty.

\bibitem[{Bernheim(1984)}]{Bernheim_1984}
\textsc{Bernheim, D.B.} (1984).
\newblock Rationalizable Strategic Behavior.
\newblock \emph{Econometrica}, \textbf{52}, 1007--1028.

\bibitem[{Bonanno(2008)}]{Bonanno_2008}
\textsc{Bonanno, G.} (2008).
\newblock A Syntactic Approach to Rationality in Games with Ordinal Payoffs.
\newblock In \emph{Logic and the Foundations of Game and Decision Theory (LOFT
  7)} (G.~Bonanno, W.~van~der Hoek, M.~Wooldridge, editors). Amsterdam
  University Press, (pp. 59--86).

\bibitem[{Bonanno(2015)}]{Bonanno_2015}
---{}---{}--- (2015).
\newblock Epistemic Foundations of Game Theory.
\newblock In \emph{Handbook of Epistemic Logic} (J.Y. Halpern, B.~Kooi,
  W.~van~der Hoek, H.~van Ditmarsch, editors). College Publications, London.

\bibitem[{Bonanno(2018)}]{Bonanno_2018}
---{}---{}--- (2018).
\newblock \emph{Game Theory}.
\newblock 2nd edition.

\bibitem[{Bonanno \& Nehring(1998)}]{Bonanno_Nehring_1998}
\textsc{Bonanno, G., Nehring, K.} (1998).
\newblock On Stalnaker's notion of strong rationalizability and Nash
  equilibrium in perfect information games.
\newblock \emph{Theory and Decision}, \textbf{45}(3), 291--295.

\bibitem[{Bonanno \& Tsakas(2018)}]{Bonanno_Tsakas_2018}
\textsc{Bonanno, G., Tsakas, E.} (2018).
\newblock Common Belief of Weak-Dominance Rationality in Strategic-Form Games:
  A Qualitative Analysis.
\newblock \emph{Games and Economic Behavior}, \textbf{112}, 231--241.

\bibitem[{Borel(1921)}]{Borel_1921}
\textsc{Borel, {\'E}.} (1921).
\newblock La Th{\'e}orie du Jeu et les {\'E}quations Int{\'e}grales {\`a} Noyau
  Sym{\'e}trique Gauche.
\newblock \emph{Comptes Rendus Hebdomadaires des S{\'e}ances de l'Acad{\'e}mie
  des Sciences}, \textbf{173}, 1304--1308.

\bibitem[{Borel(1927)}]{Borel_1927}
---{}---{}--- (1927).
\newblock Sur les Syst{\`e}mes de Formes Lin{\'e}aires {\`a} D{\'e}terminant
  Sym{\'e}trique Gauche et sur la Th{\'e}orie G{\'e}n{\'e}rale des Jeux.
\newblock \emph{Comptes Rendus Hebdomadaires des S{\'e}ances de l'Acad{\'e}mie
  des Sciences}, \textbf{186}, 52--54.

\bibitem[{B\"orgers(1993)}]{Borgers_1993}
\textsc{B\"orgers, T.} (1993).
\newblock Pure Strategy Dominance.
\newblock \emph{Econometrica}, \textbf{61}, 423--430.

\bibitem[{Brandenburger(2003)}]{Brandenburger_2003}
\textsc{Brandenburger, A.} (2003).
\newblock On the Existence of a `Complete' Possibility Structure.
\newblock In \emph{Cognitive Processes and Economic Behavior} (M.~Basili,
  N.~Dimitri, I.~Gilboa, editors). Routledge, (pp. 30--34).

\bibitem[{Brandenburger \& Dekel(1987)}]{Brandenburger_Dekel_1987}
\textsc{Brandenburger, A., Dekel, E.} (1987).
\newblock Rationalizability and correlated equilibria.
\newblock \emph{Econometrica: Journal of the Econometric Society}, 1391--1402.

\bibitem[{Brandenburger et~al.(2008)Brandenburger, Friedenberg, \&
  Keisler}]{Brandenburger_et_al_2008}
\textsc{Brandenburger, A., Friedenberg, A., Keisler, J.H.} (2008).
\newblock Admissibility in Games.
\newblock \emph{Econometrica}, \textbf{76}, 307--352.

\bibitem[{Brandenburger \& Keisler(2006)}]{Brandenburger_Keisler_2006}
\textsc{Brandenburger, A., Keisler, J.H.} (2006).
\newblock An Impossibility Theorem on Beliefs in Games.
\newblock \emph{Studia Logica: An International Journal for Symbolic Logic},
  \textbf{84}, 211--240.

\bibitem[{Brunner et~al.(2021)Brunner, Kauffeldt, \& Rau}]{Brunner_et_al_2021}
\textsc{Brunner, C., Kauffeldt, T.F., Rau, H.} (2021).
\newblock Does Mutual Knowledge of Preferences Lead to More Nash Equilibrium
  Play? Experimental Evidence.
\newblock \emph{European Economic Review}, \textbf{135}, 103735.

\bibitem[{Chen \& Micali(2015)}]{Chen_Micali_2015}
\textsc{Chen, J., Micali, S.} (2015).
\newblock Mechanism Design with Possibilistic Beliefs.
\newblock \emph{Journal of Economic Theory}, \textbf{156}, 77--102.

\bibitem[{Chen et~al.(2015{\natexlab{a}})Chen, Micali, \&
  Pass}]{Chen_et_al_2015a}
\textsc{Chen, J., Micali, S., Pass, R.} (2015{\natexlab{a}}).
\newblock {\GG{20151}} Tight Revenue Bounds with Possibilistic Beliefs and
  Level-$k$ Rationality.
\newblock \emph{Econometrica}, \textbf{83}, 1619--1639.

\bibitem[{Chen et~al.(2015{\natexlab{b}})Chen, Micali, \&
  Pass}]{Chen_et_al_2015b}
---{}---{}--- (2015{\natexlab{b}}).
\newblock {\GG{20152}}Supplement to ``Tight Revenue Bounds with Possibilistic
  Beliefs and Level-$k$ Rationality".
\newblock \emph{Econometrica}, \textbf{83}, 1619--1639.

\bibitem[{Dekel \& Siniscalchi(2015)}]{Dekel_Siniscalchi_2015}
\textsc{Dekel, E., Siniscalchi, M.} (2015).
\newblock Epistemic Game Theory.
\newblock In \emph{Handbook of Game Theory} (H.P. Young, S.~Zamir, editors),
  volume~IV. North-Holland, Amsterdam.

\bibitem[{Dominiak \& Guerdjikova(2021)}]{Dominiak_Guerdjikova_2021}
\textsc{Dominiak, A., Guerdjikova, A.} (2021).
\newblock Pessimism and Optimism towards New Discoveries.
\newblock \emph{Theory and Decision}, \textbf{90}, 321--370.

\bibitem[{Dominiak \& Schipper(2019)}]{Dominiak_Schipper_2019}
\textsc{Dominiak, A., Schipper, B.C.} (2019).
\newblock Common Belief in Choquet Rationality and Ambiguity Attitudes --
  Extended Abstract.
\newblock \emph{Proceedings Seventeenth Conference on Theoretical Aspects of
  Rationality and Knowledge (TARK)}.

\bibitem[{Eichberger \& Kelsey(2014)}]{Eichberger_Kelsey_2014}
\textsc{Eichberger, J., Kelsey, D.} (2014).
\newblock Optimism and Pessimism in Games.
\newblock \emph{International Economic Review}, \textbf{55}, 483--505.

\bibitem[{Friedenberg(2010)}]{Friedenberg_2010}
\textsc{Friedenberg, A.} (2010).
\newblock When Do Type Structures Contain All Hierarchies of Beliefs?
\newblock \emph{Games and Economic Behavior}, \textbf{68}, 108--129.

\bibitem[{Friedenberg(2019)}]{Friedenberg_2019}
---{}---{}--- (2019).
\newblock Bargaining under Strategic Uncertainty: the Role of Second-Order
  Optimism.
\newblock \emph{Econometrica}, \textbf{87}, 1835--1865.

\bibitem[{Friedenberg \& Keisler(2021)}]{Friedenberg_Keisler_2021}
\textsc{Friedenberg, A., Keisler, H.J.} (2021).
\newblock Iterated Dominance Revisited.
\newblock \emph{Economic Theory}, \textbf{72}, 377--421.

\bibitem[{Gilboa \& Marinacci(2011)}]{Gilboa_Marinacci_2011}
\textsc{Gilboa, I., Marinacci, M.} (2011).
\newblock Ambiguity and the Bayesian Paradigm.
\newblock Mimeo.

\bibitem[{Gilboa \& Schmeidler(1989)}]{Gilboa_Schmeidler_1989}
\textsc{Gilboa, I., Schmeidler, D.} (1989).
\newblock Maxmin Expected Utility with a Non-Unique Prior.
\newblock \emph{Journal of Mathematical Economics}, \textbf{18}, 141--153.

\bibitem[{Gossner \& Kuzmics(2019)}]{Gossner_Kuzmics_2019}
\textsc{Gossner, O., Kuzmics, C.} (2019).
\newblock Preferences under Ignorance.
\newblock \emph{International Economic Review}, \textbf{60}, 241--257.

\bibitem[{Guarino \& Tsakas(2021)}]{Guarino_Tsakas_2021}
\textsc{Guarino, P., Tsakas, E.} (2021).
\newblock Common Priors under Endogenous Uncertainty.
\newblock \emph{Journal of Economic Theory}, \textbf{194}, 105254.

\bibitem[{Guo \& Yannelis(2021)}]{Guo_Yannelis_2021}
\textsc{Guo, H., Yannelis, N.C.} (2021).
\newblock Full Implementation under Ambiguity.
\newblock \emph{American Economic Journal: Microeconomics}, \textbf{13},
  148--178.

\bibitem[{Hillas \& Samet(2014)}]{Hillas_Samet_2014}
\textsc{Hillas, J., Samet, D.} (2014).
\newblock Weak Dominance: A Mistery Cracked.
\newblock Mimeo.

\bibitem[{Jakobsen(2020)}]{Jakobsen_2020}
\textsc{Jakobsen, A.M.} (2020).
\newblock A Model of Complex Contracts.
\newblock \emph{American Economic Review}, \textbf{110}, 1243--1273.

\bibitem[{Kelsey \& Quiggin(1992)}]{Kelsey_Quiggin_1992}
\textsc{Kelsey, D., Quiggin, J.} (1992).
\newblock Theories of Choice under Ignorance and Uncertainty.
\newblock \emph{Journal of Economic Surveys}, \textbf{6}, 133--153.

\bibitem[{Luce \& Raiffa(1957)}]{Luce_Raiffa_1957}
\textsc{Luce, D.R., Raiffa, H.} (1957).
\newblock \emph{Games and Decisions. Introduction and Critical Survey}.
\newblock Wiley, New York.

\bibitem[{Mariotti(2003)}]{Mariotti_2003}
\textsc{Mariotti, T.} (2003).
\newblock Hierarchies of Compact Beliefs and Rationalizable Behavior.
\newblock \emph{Economics Letters}, \textbf{79}, 199--204.

\bibitem[{Mariotti et~al.(2005)Mariotti, Meier, \&
  Piccione}]{Mariotti_et_al_2005}
\textsc{Mariotti, T., Meier, M., Piccione, M.} (2005).
\newblock Hierarchies of Beliefs for Compact Possibility Models.
\newblock \emph{Journal of Mathematical Economics}, \textbf{41}, 303--324.

\bibitem[{Milnor(1954)}]{Milnor_1954}
\textsc{Milnor, J.} (1954).
\newblock Games against Nature.
\newblock In \emph{Decision Processes} (R.M. Thrall, C.H. Coombs, R.L. Davis,
  editors). John Wiley \& Sons, Inc.

\bibitem[{Morris(1997)}]{Morris_1997}
\textsc{Morris, S.E.} (1997).
\newblock Alternative Definitions of Knowledge.
\newblock In \emph{Epistemic Logic and the Theory of Games and Decisions}
  (M.~Bacharach, L.~G{\'e}rard-Varet, P.~Mongin, H.~Shin, editors). Kluwer
  Academic.

\bibitem[{Nash(1950)}]{Nash_1950}
\textsc{Nash, J.F.} (1950).
\newblock Equilibrium Points in $n$-Person Games.
\newblock \emph{Proceedings of the National Academy of Sciences}, \textbf{36},
  48--49.

\bibitem[{Nash(1951)}]{Nash_1951}
---{}---{}--- (1951).
\newblock Non-Cooperative Games.
\newblock \emph{Annals of Mathematics}, 286--295.

\bibitem[{Nikzad(2021)}]{Nikzad_2021}
\textsc{Nikzad, A.} (2021).
\newblock Persuading a Pessimist: Simplicity and Robustness.
\newblock \emph{Games and Economic Behavior}, \textbf{129}, 144--157.

\bibitem[{Osborne \& Rubinstein(1994)}]{Osborne_Rubinstein_1994}
\textsc{Osborne, M.J., Rubinstein, A.} (1994).
\newblock \emph{A Course in Game Theory}.
\newblock MIT Press.

\bibitem[{Pearce(1984)}]{Pearce_1984}
\textsc{Pearce, D.G.} (1984).
\newblock Rationalizable Strategic Behavior and the Problem of Perfection.
\newblock \emph{Econometrica}, \textbf{52}, 1029--1050.

\bibitem[{Perea(2012)}]{Perea_2012}
\textsc{Perea, A.} (2012).
\newblock \emph{Epistemic Game Theory: Reasoning and Choice}.
\newblock Cambridge University Press.

\bibitem[{Samet(2010)}]{Samet_2010}
\textsc{Samet, D.} (2010).
\newblock Agreeing to Disagree: The Non-Probabilistic Case.
\newblock \emph{Games and Economic Behavior}, \textbf{69}, 169--174.

\bibitem[{Samet(2013)}]{Samet_2013}
---{}---{}--- (2013).
\newblock Common Belief of Rationality in Games of Perfect Information.
\newblock \emph{Games and Economic Behavior}, \textbf{79}, 192--200.

\bibitem[{Samuelson(1992)}]{Samuelson_1992}
\textsc{Samuelson, L.} (1992).
\newblock Dominated Strategies and Common Knowledge.
\newblock \emph{Games and Economic Behavior}, \textbf{4}, 284--313.

\bibitem[{Savage(1954)}]{Savage_1954}
\textsc{Savage, L.J.} (1954).
\newblock \emph{The Foundations of Statistics}.
\newblock Wiley, New York.

\bibitem[{Schipper(2021)}]{Schipper_2021}
\textsc{Schipper, B.C.} (2021).
\newblock The Evolutionary Stability of Optimism, Pessimism, and Complete
  Ignorance.
\newblock \emph{Theory and Decision}, \textbf{90}, 417--454.

\bibitem[{Schmeidler(1989)}]{Schmeidler_1989}
\textsc{Schmeidler, D.} (1989).
\newblock Subjective Probability and Expected Utility without Additivity.
\newblock \emph{Econometrica}, \textbf{57}, 571--587.

\bibitem[{Siniscalchi(2008)}]{Siniscalchi_2008}
\textsc{Siniscalchi, M.} (2008).
\newblock ``Epistemic Game Theory: Beliefs and Types''.
\newblock In \emph{The New Palgrave Dictionary of Economics} (S.N. Durlauf,
  L.E. Blume, editors), 2nd edition. Palgrave Macmillan, New York.

\bibitem[{Stalnaker(1994)}]{Stalnaker_1994}
\textsc{Stalnaker, R.} (1994).
\newblock On the Evaluation of Solution Concepts of Games.
\newblock \emph{Theory and Decision}, \textbf{37}, 49--74.

\bibitem[{Stalnaker(1998)}]{Stalnaker_1998}
---{}---{}--- (1998).
\newblock Belief Revision in Games: Forward and Backward Induction.
\newblock \emph{Mathematical Social Sciences}, \textbf{36}, 31--56.

\bibitem[{Tan \& da~Costa~Werlang(1988)}]{Tan_daCosta_1988}
\textsc{Tan, T.C.C., da~Costa~Werlang, S.R.} (1988).
\newblock The Bayesian Foundations of Solution Concepts of Games.
\newblock \emph{Journal of Economic Theory}, \textbf{45}, 370--391.

\bibitem[{von Neumann(1928)}]{vonNeumann_1928}
\textsc{von Neumann, J.} (1928).
\newblock Zur Theorie des Gesellschaftspiele.
\newblock \emph{Mathematische Annalen}, \textbf{100}, 295--320.

\bibitem[{von Neumann \& Morgenstern(1944)}]{vonNeumann_Morgenstern_1944}
\textsc{von Neumann, J., Morgenstern, O.} (1944).
\newblock \emph{Theory of Games and Economic Behavior}.
\newblock Princeton University Press, Princeton.

\bibitem[{Wald(1950)}]{Wald_1950}
\textsc{Wald, A.} (1950).
\newblock \emph{Statistical Decision Functions}.
\newblock John Wiley and Sons.

\bibitem[{Weinstein(2016)}]{Weinstein_2016}
\textsc{Weinstein, J.} (2016).
\newblock The Effect of Changes in Risk Attitude on Strategic Behavior.
\newblock \emph{Econometrica}, \textbf{84}, 1881--1902.

\bibitem[{Wilson(1987)}]{Wilson_1987}
\textsc{Wilson, R.B.} (1987).
\newblock Game-Theoretic Analyses of Trading Processes.
\newblock In \emph{Advances in Economics and Econometrics, 5th World Congress
  of the Econometric Society} (T.~Bewley, editor). Cambridge University Press,
  Cambridge.

\bibitem[{Yildiz(2007)}]{Yildiz_2007}
\textsc{Yildiz, M.} (2007).
\newblock Wishful Thinking in Strategic Environments.
\newblock \emph{Review of Economic Studies}, \textbf{74}, 319--344.

\bibitem[{Zamir(2009)}]{Zamir_2009}
\textsc{Zamir, S.} (2009).
\newblock ``Bayesian Games: Games with Incomplete Information''.
\newblock In \emph{Encyclopedia of Complexity and Systems Science} (R.A.
  Meyers, editor). Springer.

\bibitem[{Ziegler \& Zuazo-Garin(2020)}]{Ziegler_Zuazo-Garin_2020}
\textsc{Ziegler, G., Zuazo-Garin, P.} (2020).
\newblock Strategic Cautiousness as an Expression of Robustness to Ambiguity.
\newblock \emph{Games and Economic Behavior}, \textbf{119}, 197--215.

\end{thebibliography}

\bibliographystyle{ampersand_standard_capital_natbib}

\end{document}